\documentclass{article}
\usepackage{authblk}
\usepackage{lscape}
\usepackage{graphicx} 
\usepackage{comment}
\usepackage{siunitx}
\graphicspath{ {./Figures/} }
\usepackage{xcolor}
\usepackage{soul}
\usepackage{hyperref}
\usepackage{multirow} 

\begin{document}

\title{ConCeG: Contextual Cellular Growth for Realistic Multicellular Grey Matter Phantoms in Diffusion MRI}

\author[1,2]{Charlie Aird-Rossiter}
\author[1,2]{Kadir \c{S}im\c{s}ek}
\author[1,2]{Ma{\"e}liss Jallais}
\author[1]{Derek K. Jones}
\author[3]{Lida Kanari}
\author[1,2]{Marco Palombo}

\affil[1]{Cardiff University Brain Research Imaging Centre (CUBRIC), School of Psychology, Cardiff University, Cardiff, United Kingdom}
\affil[2]{School of Computer Science and Informatics, Cardiff University, Cardiff, United Kingdom}
\affil[3]{Department of Mathematics, University of Oxford, Oxford, UK}

\date{}                     

\maketitle

\begin{abstract}
Accurate interpretation of diffusion magnetic resonance imaging (dMRI) signals in grey matter (GM) remains challenging due to the complex, heterogeneous, and densely packed cellular environment. 
Numerical phantoms provide a controlled framework for investigating the relationship between microstructure and diffusion signals, yet existing approaches often lack the morphological realism and multi-cellular organisation required to faithfully represent GM tissue. 

In this work, we introduce Contextual Cellular Growth (ConCeG), a generative framework for creating individual cells or constructing dense, three-dimensional, multi-cellular GM substrates informed by real neuronal and glial morphologies. The method combines topological neuron synthesis with a spatially constrained growth network, allowing for the controlled generation of heterogeneous cellular environments with realistic intra- and extracellular compartments. 
Synthetic cells are generated using morphological and topological characteristics derived from biological reconstructions.

We validate the framework through comparisons of structural features with real cellular data, demonstrating strong agreement in branch order, length, and angle distributions.
Power spectrum analysis further shows that  intracellular compartments reproduce the spatial correlations observed in biological tissue. 
Simulation of diffusion signals within ConCeG generated substrates and subsequent SANDI parameter estimates demonstrate ConCeGs compatibility with Monte Carlo simulators.
Together, these results show ConCeG provides a biologically grounded framework for generating grey matter substrates suitable for large scale diffusion MRI simulation.
\end{abstract}

\section{Introduction}

Grey matter (GM) is composed of a densely packed and highly heterogeneous arrangement of neural cells. 
These cellular structures display a highly diverse morphology, with each cell type tailored to its functional role \cite{Miller1984,levitan2002neuron,Zeng2017}, from large, planar, highly branched Purkinje neurons, to pyramidal neurons with prominent apical dendrites, and smaller, more isotropic glial cells such as astrocytes and microglia. 

Throughout brain development and aging, significant changes occur in GM microstructure, including apparent reductions in cell density \cite{Lee2022}, dendritic complexity \cite{Dickstein2013,Dickstein2007}, and soma size \cite{Nassif2022}. 
These changes contribute to cortical thinning with age and are accelerated in patients with Alzheimer's \cite{Dickstein2013}.

In addition, many neurological and neurodegenerative disorders are characterized by changes in GM microstructure, including neuronal loss \cite{Surmeier2017,ross2011huntington}, alterations in dendritic structure, manifesting as reductions in branch size and complexity \cite{Baloyannis2009,Kweon2017}, atypical composition, and the presence of abnormal neural cells \cite{castano2025architecture}. 

As such, there is strong motivation to develop methods capable of characterising GM cellular structure and composition \textit{in vivo} and non-invasively.

Diffusion magnetic resonance imaging (dMRI) is a powerful non-invasive technique for probing brain microstructure. 
dMRI sensitises the MR signal to the displacement of water at the cellular level within tissue, using  magnetic field gradients \cite{callaghan1993principles}. 
The observed signal attenuation depends on both the applied gradients and the displacement of water molecules during encoding. 

Since water undergoes Brownian motion, its displacement is governed by the underlying tissue microstructure, which hinders and restricts diffusion \cite{stejskal1965spin,Bihan1995}. 
However, the relationship between signal attenuation and tissue microstructure is indirect and requires biophysical modelling to estimate parameters reflecting the underlying microstructure \cite{Alexander2019, novikov2019quantifying}.

A range of biophysical models have been proposed to estimate microstructural features from diffusion signals, including neurite orientation dispersion \cite{zhang2012noddi, coelho2022reproducibility} and axon diameter \cite{Assaf2008, veraart2020noninvasive}. 

Numerical simulations play a crucial role in evaluating these models, enabling controlled validation of accuracy and robustness \cite{Hall2009,alexander2010orientationally,Fieremans2010,Rafael-Patino2020,lee2021realistic}. 
Numerical phantoms are particularly valuable as they provide control over microstructural parameters and a known ground truth.

Despite their advantages, creating realistic numerical phantoms remains challenging, with many studies using simplified geometries such as parallel cylinders \cite{Fieremans2010,Nilsson2009,Nilsson2010, alexander2010orientationally, Hall2009}, to approximate white matter (WM) fibre bundles. 
Subsequent work introduced more complex geometries including beading \cite{budde2010neurite,landman2010complex}, and undulation \cite{Brabec2020,Nilsson2012}, highlighting the importance of microstructural realism. 
Whilst these simple geometries provide insight into the diffusion dynamics they drastically reduce the complexity of the tissue they aim to represent, limiting the geometric complexity to a small set of microstructural features.
In reality, neurites (axons and dendrites) exhibit high levels of complexity, as revealed by electron microscopy (EM) \cite{abdellah2018neuromorphovis}.

One approach is to conduct simulations within these real geometries, reconstructed from three-dimensional EM \cite{lee2020time,Lee2020,abdollahzadeh2025scattering}, to truly capture the tissue complexity. 
While these methods provide realistic tissue simulations, they are limited in size, with EM volumes generally being small (on the order of $10^3$ $\mu m^3$). Additionally, it is not possible to fully “tune” the phantoms to probe individual microstructural features (i.e., vary as arbitrarily as possible selected features).

As such, there has been recent approaches that aim to generate realistic three-dimensional digital (or numerical) phantoms that better reflect the complexity of real tissues, incorporating multiple microstructural features in a fully tuneable way \cite{Callaghan2020,Ginsburger2018,Villarreal-Haro2023,Winther2024,nguyen2026caterpillar,Abdollahzadeh2025}, allowing for the generation of phantoms with undulating, beading, orientations dispersion, and fibre crossing.

These approaches have primarily focused on WM, where simplifying assumptions can be made, such as treating axons as non-exchanging cylinders \cite{Jelescu2017}. However, GM microstructure presents a greater challenge due to its more heterogeneous and multi-cellular structure \cite{jelescu2020challenges}. 



To better understand how specific morphological features influence the diffusion signal, there is a need for realistic numerical phantoms tailored to GM microstructure, similar to those developed for WM. 
Some methods partially address this need. 
For example, WM digital phantoms generators such as MEDUSA \cite{Ginsburger2019} and Caterpillar \cite{nguyen2026caterpillar} incorporate glial cells into their substrates, while other approaches focus solely on simulating the intracellular signal using either microscopy-derived meshes \cite{kiselev2026does,Fang2020} or synthetically generated cellular structures \cite{palombo2019generative,ianus2021mapping}.

However, despite these advances, there remains a lack of a dedicated GM phantom generator that captures the full complexity of cellular morphology and tissue organisation, as well as the complex GM extracellular space.

\begin{figure}[ht!]
    \centering
    \includegraphics[scale=0.75]{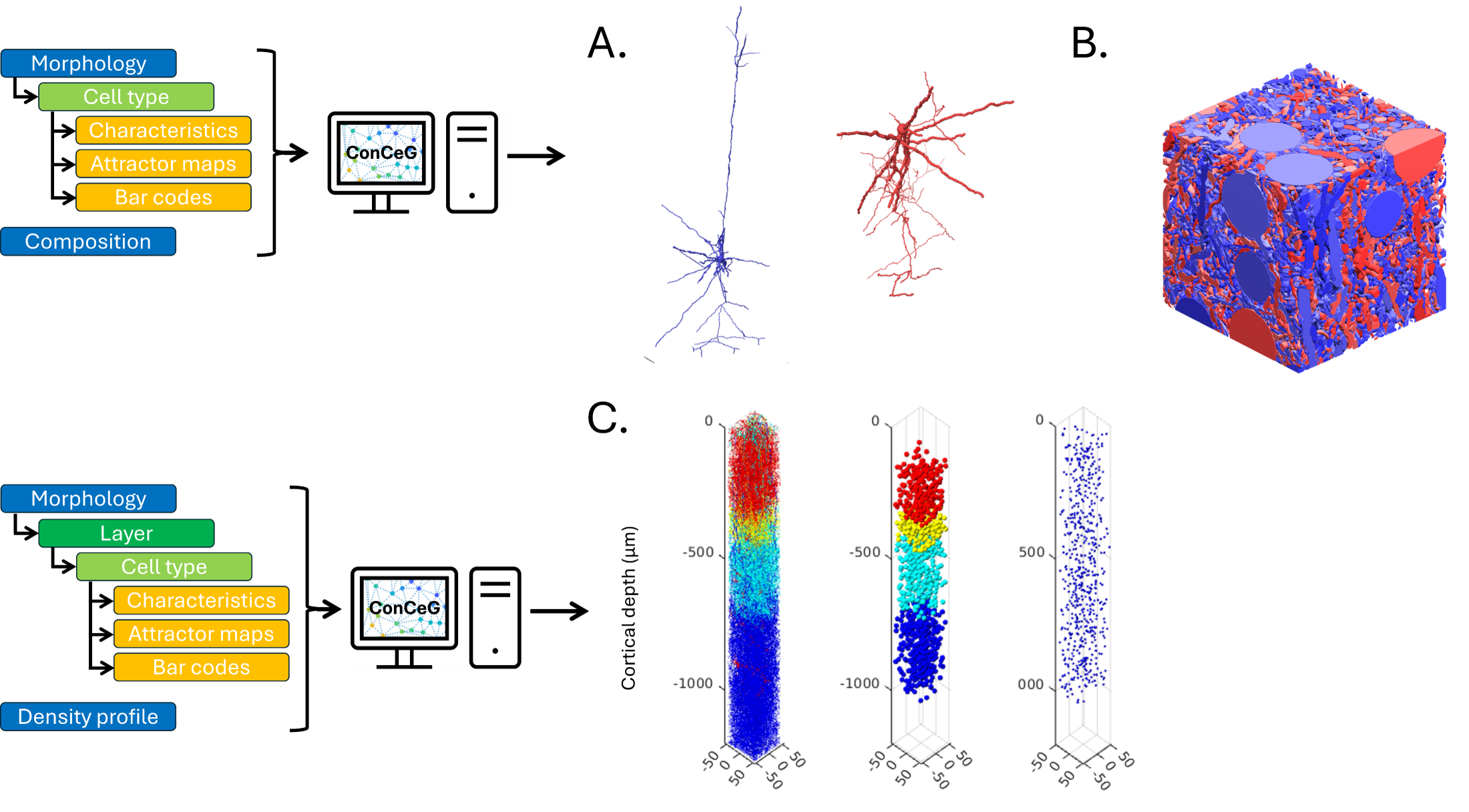}
    \caption{\textbf{Visualisation of ConCeG output.} ConCeG uses a morphological dictionary to generate individual neuronal morphologies (A, example pyramidal (blue) and basket (red) cell) and user-defined cell compositions to generate heterogeneous substrates (B, a substrate containing pyramidal (blue) and basket (red) cells at a 4:1 ratio). Combining a cortical layer-specific morphological dictionary with a soma density profile enables the generation of cortical grey matter columns (C, from left to right: the full cortical column, neuronal soma placement, and glial soma placement).  }
    \label{fig:ConCeGOut}
\end{figure}

The aim of this work is to address this unmet need. We present ConCeG, a framework for \textbf{Con}textual \textbf{Ce}llular \textbf{G}rowth, that enables controlled generation of dense, heterogeneous, multi-cellular GM substrates compatible with Monte Carlo diffusion simulations, see Figure \ref{fig:ConCeGOut}.

The paper is organised as follows. We first describe the implementation of ConCeG and illustrate how it can be used to generate: replicas of individual cell types (Figure \ref{fig:ConCeGOut}A); volumes with arbitrary cell-type mixtures (Figure \ref{fig:ConCeGOut}B); and complete mesoscopic cortical columns (Figure \ref{fig:ConCeGOut}C). We then present a series of validation experiments assessing the realism and morphological accuracy of ConCeG-generated tissues. Finally, we demonstrate the potential of the framework through exemplar diffusion MRI simulations as a proof of concept.  

\section{Methods}

\subsection{Overview}
The ConCeG pipeline consists of:
\begin{itemize}
    \item Morphological characterisation
    \item Voxel initialisation
    \item Synthetic growth
    \item Mesh generation
\end{itemize}

\subsection{Morphological Characterisation}
\begin{figure}[hbt!]
    \centering
    \includegraphics[scale=0.75]{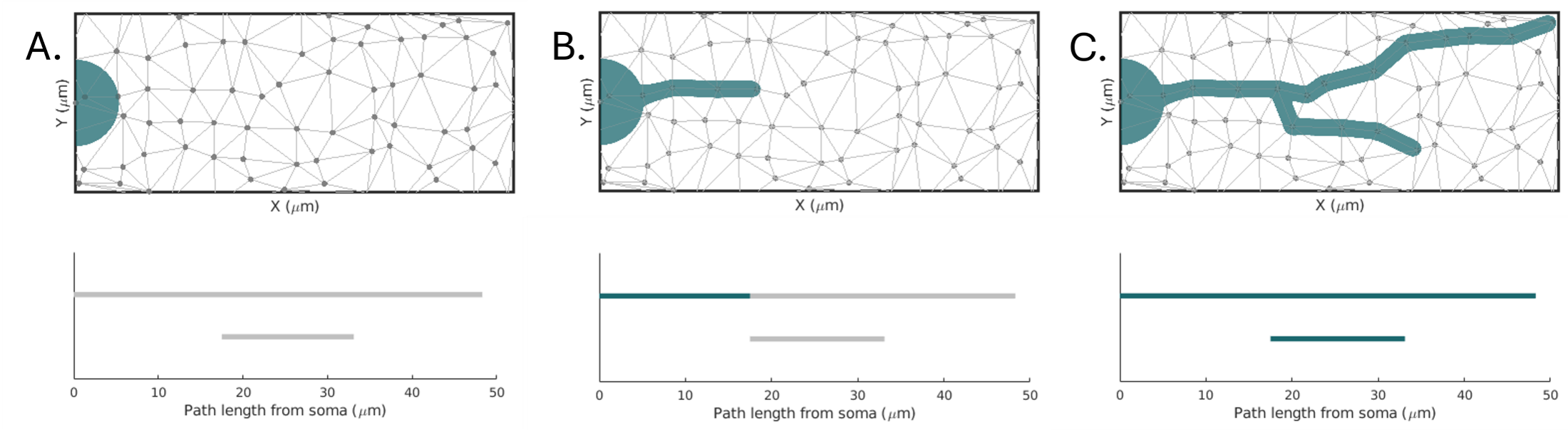}
    \caption{\textbf{Visualisation of topologically informed branching}, 
    A. Initial network configuration, with a soma on the left and the bar code for the projection is shown below. 
    B. Initial growth, as the projection navigates the network its path length increases, increasing the probability of initiating the second branch seen in the bar code (when the current path length is equal to the initiation length of a branch the probability is equal to one). 
    C. Complete growth, when projections have achieved their prescribed termination as defined by the bar code the projection is complete.}
    \label{fig:branching}
\end{figure}

To accurately recreate neural structures, it is first necessary to characterise their morphology. Neuronal reconstructions are first obtained from open repositories (NeuroMorpho.org \cite{ascoli2007neuromorpho} and the Allen Brain Atlas \cite{peng2021morphological}) and analysed for their structural, spatial, and topological properties, as described in more details in \cite{aird2026decoding}.
Briefly, all reconstructions are pre-processed to remove dendritic spines, as their inclusion obscures the underlying branching structure, drastically overestimating branch order. 
For pyramidal neurons, apical and basal dendrites are analysed separately due to their distinct morphologies.
Structural parameters were first calculated, including branch order and branch angle distributions. These distributions inform the branching behaviour of synthetic cells during the contextual growth, as well as mean dendritic branch diameter and coefficient of variation and soma size.
To capture spatial distribution of dendritic branches, terminal points from each reconstruction are identified and projected onto a unit sphere centred at the soma. These projected points define a directional distribution used to assign what we call "attractor points", i.e. target points towards which each synthetic cellular projection grows, loosely mimicking the influence of chemical gradients during biological growth.
Finally, the topological persistence barcode \cite{kanari2018topological} is computed for each cell using path length from the soma as the persistence parameter. This barcode provides a concise representation of the branching structure and is used to inform branch initiation and termination during synthetic growth (see Figure \ref{fig:branching}). Corresponding branch diameters were also recorded to accurately reproduce dendritic tapering which is not well defined from reconstructions obtained from microscopy.

\subsection{Network Initialisation}
Soma positions are first placed within a three-dimensional domain, i.e. a digital voxel, according to a user-defined soma density distribution and specified cell-type composition. Each soma defines the origin of a synthetic cell and is assigned a cell-type-specific topological barcode describing its dendritic projections.
Following the same growth strategy as in ConFiG \cite{Callaghan2020}, the digital voxel space is then populated with randomly distributed nodes representing allowable growth locations.
To allow for periodic boundaries, nodes are placed on the faces defining the volume and mirrored on opposing sides.
Each node is assigned a maximum allowable dendritic radius, encoding spatial occupancy constraints within the local space.
All nodes were connected via a three-dimensional Delaunay triangulation. The edges of the resulting graph define the allowable steps that cellular projections may take during growth.

\subsection{Synthetic Growth}
Cellular projections are generated by iteratively extending every segment through the network according to biologically informed cost functions.
Each projection is assigned an attractor point sampled from the given attractor map. Projections grow toward their assigned attractor following the path that minimizes the assigned cost function, inherently and simultaneously avoiding existing cellular structures within the voxel space (this is further explained in the following paragraph). Growth proceeds until the prescribed terminal length, as defined by the persistence barcode, is achieved.
The network’s maximum allowable radius constraints are updated as growth progresses to reflect the current occupancy of the voxel space. 
Branching behavior is governed by the assigned persistence barcode. At each growth step, the probability of initiating a secondary branch is evaluated according to the barcode structure. When branching is triggered, a new projection is initiated, assigned its own attractor point, and grows using the same biological cost functions.
This process is repeated for all barcode intervals associated with each soma and for all cells within the voxel domain. Growth is performed sequentially, with one projection grown at a time per cell, ensuring no single cell dominates through unrestricted growth, until all cellular projections are grown.
The resulting cellular morphologies are saved in SWC format, the standard form for reporting and handling real cellular reconstructions.

\subsection{Network Navigation}
Network navigation is governed by biologically motivated cost functions that mimic the chemical cues influencing dendritic growth. 
Whilst the growth and arborisation of neural cells is highly complex and defined by numerable molecular and electrical signals, a drastically simplified motivation is employed here. 
The functions governing cellular projections growth are reduced to chemoattraction and arbor collapse which are described in more detail bellow. 

As projections grow through the voxel space, the direction of growth is determined by the attractor and other structures within the space, loosely mimicking chemoattraction. From the projection’s current position at node $s$, candidate nodes $c$ are identified. Each node has an associated maximum allowable projection radius $r_{c}$, defining the largest radius that can occupy that location.
Two terms define the growth cost:

\begin{itemize}
    \item$l_{t}$: a directional term encouraging movement toward the attractor point, $t$, while penalising excessively large steps. Defined as,
    \begin{equation}
    l_t = \frac{1}{2} \cdot \frac{|s-c|}{1+|s-c|} \cdot \left(1 - \frac{(c-s)\cdot(t-s)}{|c-s||t-s|}\right)
    \end{equation}
    \item $l_{r}$: a constraint term penalising transitions that would require projection constriction beyond the allowable radius at that node. Defined as, 
    \begin{equation}
    l_r = \max\left(0, \frac{1}{r_0}(r_0 - r_c)\right)
    \end{equation}
    where $r_{0}$ is the desired dendrite radius.
\end{itemize}

The total cost for each node is given by:  
\begin{equation}
l = l_{t} + fl_{r}
\end{equation}

Where $f$ is a weighting factor between the two terms, allowing for growth to be preferentially determined by attractor targeting or preservation of projection radius.  
The candidate node with the lowest cost is selected, and the projection advances to that node. This process continues iteratively until the prescribed termination length is reached or no viable candidate nodes remain.

Due to the static nature of the underlying network and the increasing occupancy of the voxel space during growth, projections may become locally trapped.
To mitigate this, a fibre collapse mechanism was implemented. When a projection becomes trapped, it is allowed to backtrack a predefined distance $g_{0}$ and regrow while incorporating information about previously unnavigable regions. If repeated trapping occurs, growth of that projection is terminated.

\subsection{Boundary conditions}
\begin{figure}[hbt!]
    \centering
    \includegraphics[scale=0.75]{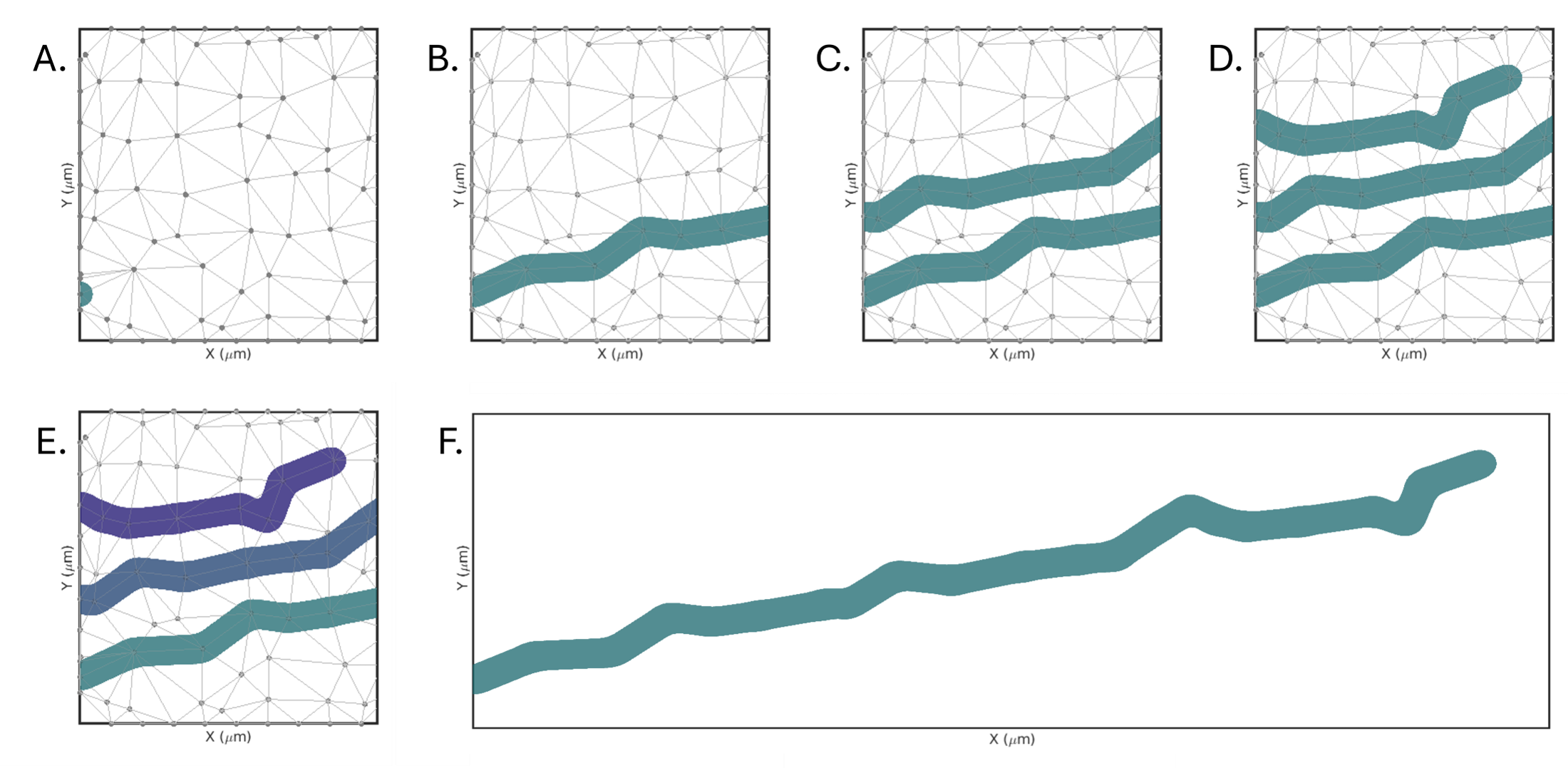}
    \caption{\textbf{Visualisation of boundary navigation}. A. shows the initial voxel configuration with a branch starting from the bottom left, growing towards an attractor that lies beyond the voxel space. B.-D. depict how active branches navigate crossing boundary's. When a boundary is met the active branch re-entrees the voxel space via the corresponding mirrored boundary node. Active branches are free to cross boundaries until they reach their assigned termination length or navigation is no longer possible. The resulting branch structure can be deconstructed such that each disconnected component can be considered to be from different cells, seen in E., or reconstructed to its full extent, seen in F. .  }
    \label{fig:boundary}
\end{figure}

Due to the large spatial extent that dendritic arbors can reach, periodic boundary conditions can be implemented to prevent branches from terminating at the edges of the voxel space. Without periodic boundaries, projections encountering the domain limits will become stuck as they are no longer able to grow towards their attractor point, resulting in incomplete morphological representation and reduced packing density. 
By allowing projections that exit one boundary of the voxel space to re-enter from the opposing boundary (see Figure \ref{fig:boundary}), synthetic cells preserve their full morphological extent while remaining confined within the voxel space. This approach maintains branch morphology, improves achievable packing density, and allows for tiling of the generated cells for larger-scale simulations.

\subsection{Branching criteria}
Projection branching follows the process outlined in the topological neuron synthesis algorithm \cite{kanari2022computational}, whereby the computed persistence bar codes are used to reconstruct the cellular structure they represent. Each persistence barcode encodes the initiation and termination length of a branch as a function of path length from the soma. During synthetic growth, branches are initiated and terminated according to these intervals, so that the resulting morphology reproduces the branching hierarchy and complexity of the original cells.

Each barcode has a projection that has an initiation length of 0 corresponding to the primary projection that emanates from the soma, it is this projection that is ‘activated' to initiate the growth. As the projection navigates through the network, at each step the probability of initiating a new branch is assessed based on the current path length, $pl$, of the active projection and the initiation length, $il$, of the remaining bars in the barcode. 

\begin{equation}
P(\text{bifurcation} \mid p_l) = e^{\lambda (p_l - il_i)}
\end{equation}

Where, $\lambda$, is a free parameter defining the probability distribution, and $il_{i}$ is the initiation length of the $i^{th}$ remaining bar in the barcode.
Once the branching criteria is satisfied the corresponding bar is activated, a branching angle is drawn from the cell-type specific distribution learned from real reconstructions and an attractor point assigned to the newly activated branch. The newly activated branch grows following the same constraints. 
Branching is assessed for all active projections until all bars in the barcode have been activated.

\subsection{Global Optimisation}
To reduce the limitations of the finite and static nature of the growth network, a post-growth optimisation step was introduced. 
This step allows constricted projections (those whose radii fall below their target radii due to occupying nodes with restricted allowable diameter during growth) to recover their intended radii after the synthetic growth process.

Following an approach similar to MEDUSA, the cellular skeletons were resampled as overlapping spheres. Colliding spheres were identified and interaction forces were computed. Repulsive forces were applied to overlapping spheres, while attractive forces were applied to nearby non-overlapping spheres to encourage higher packing density. The magnitude of these forces were weighted by the corresponding sphere volumes of the colliding pairs to mimic the dynamics seen in real axonal growth, with smaller structures more flexibly growing around larger ones  \cite{andersson2020axon}.
To preserve cellular morphology, additional forces were applied between spheres belonging to the same cell to maintain their relative separation and prevent deformation. 

\subsection{Meshing}
Following growth and optimisation, cellular skeletons represented in SWC format were converted into watertight three-dimensional surface meshes compatible with Monte Carlo diffusion simulators.
Cells were reconstructed using a metaball-based procedure utilising Blender. 
Metaballs act as signed distance fields that smoothly merge (positive interaction) or repel (negative interaction) to form continuous surfaces. 
During reconstruction, the current cell was assigned positive fields, while nearby elements from other cells were assigned negative fields to enforce non-intersection and prevent overlap.
The resulting surfaces were cleaned and remeshed to ensure watertightness and decimated to reduce the mesh size and improve simulation efficiency.

\section{Experiments}

To demonstrate the potential use of ConCeG, we generated a whole cortical column. To evaluate ConCeG, we performed structural validation comparing generated cells and extracellular space with reconstructions and segmentation from microscopy. 
As an exemplar application, we show the results of Monte Carlo simulations of diffusion MRI signals in the digital cortical column. 

\subsection{Computational time} 
To quantify the time taken to generate complete surface meshes, a series of cubic substrates with side lengths, L, ranging from 20 to 100 $\mu m$ were generated. 
This range was selected to test the performance of ConCeG across increasingly large substrate sizes while maintaining consistent generation parameters.
The substrates were all grown with the same node density, $2L^{3}$, and cellular density $\num{1.4e-5}\mu m^{-1}$ to provide comparable results across the differing sizes.

As the substrate generation can be thought of as being composed of two principle parts, the synthetic growth, and the conversion of the cellular structures into surface meshes, the time for both steps was recorded for each substrate size.
The growth stage was performed on a single CPU (parallelising is not possible whilst maintaining accurate contextual information), and the meshing was performed over 5 CPUs for all substrates. 

\subsection{Generate column of cortex} 
A synthetic cortical column was constructed, with dimensions 100 by 100 by 1200 $\mu m$ layer-specific soma densities and cellular morphologies derived from real data.
Soma density distributions were obtained from histological measurements reported by Tsai et al \cite{Tsai2009} for both neurons and glia. This data provided quantitative estimates of neural density with respect to cortical depth and was used to define the spatial distribution of somas within the voxel space. 
To replicate the cytoarchitecture as faithfully as possible, layer-specific neuronal reconstructions were obtained from the Allen Brain Atlas and assessed to give layer specific morphological characterisation for each cell type. Reconstructions from the primary visual cortex (VISp) were used as morphological references because this region contained the largest number of available reconstructions. As explicit cell-type labels were not consistently provided, neurons were categorised into two principal classes based on morphology:
\begin{itemize}
    \item Pyramidal cells, identified by the presence of apical dendrites as defined in the SWC file.
    \item Basket cells (interneurons), identified by the absence of apical projections.
\end{itemize}
Due to the limited presence of axonal components in these reconstructions, only the most complete axons were assessed and characterised.

Non-neurons were represented by astrocytes, whose morphologies were obtained from NeuroMorpho.Org \cite{Ascoli2007}.
Each soma placed within the synthetic column was assigned a morphology consistent with its cortical layer and type. The corresponding morphological statistics and persistence barcodes were used to inform synthetic growth. A ratio of 4:1, pyramidal to interneuron, was used in line with estimates from real data \cite{Gouwens2019,Tsai2009}.


\begin{figure}[hbt!]
    \centering
    \includegraphics[scale=.7]{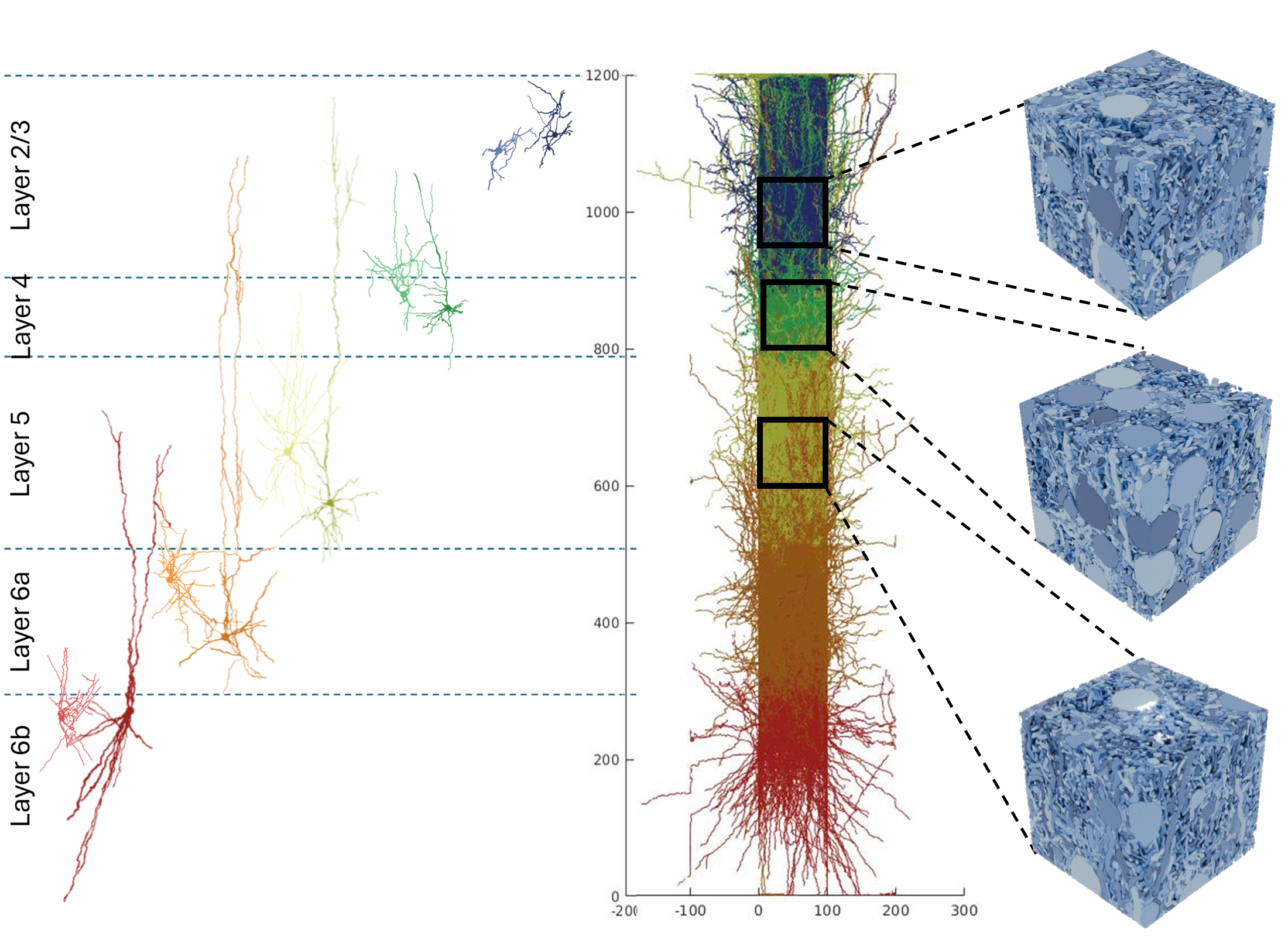}
    \caption{\textbf{Visualisation of complete cortical column}, showing exemplar cells (one basket and one pyramidal cell for each cortical layer), and meshes generated for layers 3, 4, and 5 from the synthetic column. }
    \label{fig:column}
\end{figure}

\subsection{Structural Comparison}
To assess morphological representation, synthetic cells, from the generated cortical column, were compared to their biological reference populations using classical morphometric measures.
Comparisons were performed separately for:
\begin{itemize}
    \item Basket cells
    \item Pyramidal cells
    \item Astrocytes
\end{itemize}
The following metrics were evaluated:
\begin{itemize}
    \item Total process length
    \item Branch order distribution
    \item Branch angle distribution
    \item Branch tortuosity
    \item Path length distribution
\end{itemize}
Distributions from synthetic cells were compared against the corresponding distributions from biological reconstructions to quantify agreement.

\subsection{Intracellular Power Spectrum}
A key determinant of time-dependent diffusivity in restricted geometries is the structure factor, which captures spatial correlations within cellular projections and reflects features such as tortuosity and beading.

To estimate the intracellular structure factor of synthetic cells, we followed the procedure described by A. Abdollahzadeh et al \cite{Abdollahzadeh2025}. 
Sections of dendritic branches of length 100$\mu m$ were used to find the variation in area and the resulting power spectra from the different area profiles.

The area profile was found along the branch (Figure \ref{fig:intraPower} A.), incorporating undulation and calibre variation.
Figure \ref{fig:intraPower} shows the the variation in area along the branch length with respect to the mean area $\alpha$. 
From projection area profiles the power spectrum was computed.

The structural class was derived from the resulting low frequency behaviour of the power spectrum.
The same procedure was applied to the real reconstructions used as morphological references. 
Power spectra were then compared between real and synthetic cells to assess whether the synthetic arbors reproduce the spatial frequency profiles relevant to time-dependent diffusion.

\subsection{Composition scaling}
An important structural property of grey matter is the scale-dependent organisation of its cellular composition. 
It has been shown in \cite{Ansell2024} that the volume of the largest component within subsections of ral grey matter follows a power-law relationship with sampling length scale. From which a fractal dimension $d_{f}$ of approximately 1.6 was observed across mouse, human, and fly tissue.

To determine whether the generated substrates reproduce similar powerlaw scaling of components, we measured the volume of the largest cellular component across a range of sampling scales from 2 to 75 $\mu m$. 
For each scale, cubic sub volumes were sampled from the generated substrate and the largest component within each volume was identified, see Figure \ref{fig:Composition}. 
The resulting scaling relationship between component volume and sampling length scale was then compared with that reported for \cite{Ansell2024}.

\subsection{Extracellular Characteristics}
To characterise the extracellular space, two geometric metrics are quantified, pore size distribution and tortuosity. 
Both measurements are calculated from binarised images representing the intra (0) and extracellular (1) components.

Pore size distribution is estimated by first identifying the centres of extracellular pores within the binary image. 
For each pore centre, the Euclidean distance to the nearest intracellular boundary is calculated. 
This distance represents the local pore radius and provides a measure of the available free space surrounding that point. 
Repeating this process throughout the extracellular network gives a distribution of pore sizes across the image, allowing for the quantitative assessment of extracellular pore sizes.

Tortuosity describes the allowable pathways through the extracellular space, an important metric describing transport properties of complex media. 
To calculate tortuosity, pairs of points are randomly picked within the connected extracellular region of the binary image. 
The shortest connected path between these points can then be determined using the A* algorithm \cite{dijkstra2022note}, in which neighbouring pixels are treated as connected edges within the extracellular space.
Tortuosity is defined as the ratio between Euclidean distance between the two points and the corresponding allowed path length through the extracellular space.
Values less than one indicate increasingly convoluted pathways.

This analysis was performed on binarised images from cross sections of ConCeG generated substrates and also on segmentation's openly available from the MICrONS data set \cite{microns2025functional,MICrONS2021}, obtained from electron microscopy of mouse visual cortex, to provide a comparison to real tissue.

\subsection{Diffusion MRI Signal Simulation}
To demonstrate the suitability of the generated substrates for diffusion MRI applications, Monte Carlo diffusion simulations were performed using DiSimPy \cite{Kerkel2020}. 
Simulations were carried out for both the isolated intracellular compartment, and the full substrate, containing the intracellular and extracellular compartments, using the substrates generated from the cortical column shown in Figure \ref{fig:column}. 
This allowed the diffusion-weighted signal to be evaluated both in the full tissue environment and within the intracellular space in isolation.

Diffusion-weighted signals were simulated over a range of diffusion weightings (b-values) to characterise the signal attenuation from the synthetic cellular microstructure. 
Simulations employed a pulsed-gradient spin-echo (PGSE) sequence with a diffusion time of $\Delta = 20\ ms$ and a gradient pulse duration of $\delta = 5\ ms$. 
The intrinsic diffusivity was set to $D = 2\ \mu m^{2}/ms$, and $10^{6}$ random walkers were placed within $10\ \mu m$ from the substrate boundary to avoid boundary effects and were allowed to diffuse using a simulation time step of $\tau = 10^{-3}\ ms$. Diffusion-weighted signals were calculated for b-values ranging from $0$ to $15\ ms/\mu m^{2}$, over 40 directions uniformly distributed over the unit sphere.

In addition to the diffusion-weighted signals a reduced Soma and Neurite Imaging (SANDI) model (i.e. SANDI without extracellular compartment) was fitted to the intracellular signal only, estimating the soma volume fraction, $f_{soma}$, and the effective soma radius, $r_{soma}$, and the full SANDI model was fitted to the full simulated signal.
These fits were done to assess whether the Monte Carlo simulations within the synthetic substrates produced diffusion signals that were consistent with the underlying microstructural geometry.

\section{Results}

\subsection{Computational time}
\begin{figure}[hbt!]
    \centering
    \includegraphics[scale=.42]{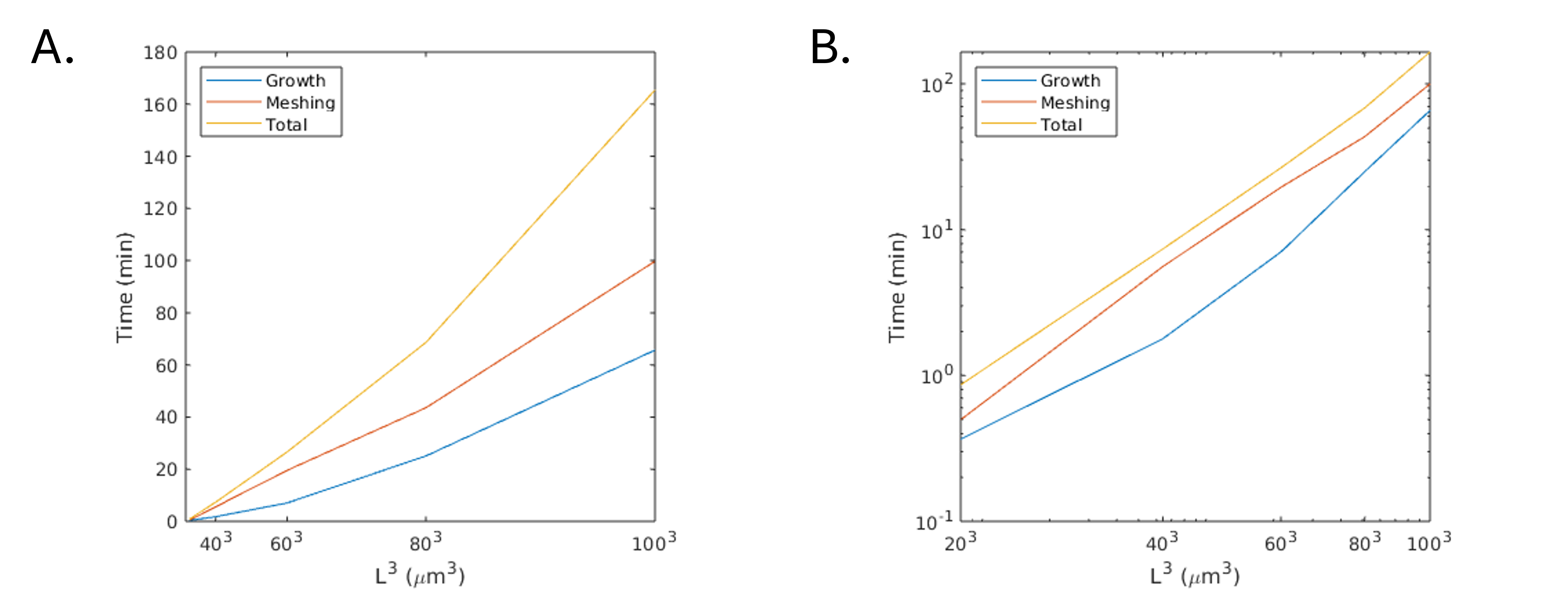}
    \caption{\textbf{Computational scaling of synthetic substrate generation.} A. Time required to generate synthetic substrates as a function of substrate volume, $L^{3}$. B. Log--log representation of generation time versus substrate volume. The fitted power-law relationship yields a scaling exponent of 1.077, indicating near-linear scaling with substrate volume.}
    \label{fig:comptime}
\end{figure}

The computational cost of substrate generation increased with substrate size (Figure \ref{fig:comptime}). 
As expected, larger substrates required longer generation times owing to the increased number of cells and nodes contained within the substrate volume.

When plotted as a function of substrate volume ($L^{3}$), generation time shows an approximately linear increase (Figure \ref{fig:comptime}A). 
To quantify this relationship, the data was additionally plotted in log-log space (Figure \ref{fig:comptime}B). 
A power-law fit yielded a scaling exponent of 1.077, indicating that computational time scales close to linearly with substrate volume.

These results indicate the growth and meshing pipeline is capable of producing meshes of $100\mu m^{3}$ in a reasonable amount of time ($~3$ hours).

\subsection{Structural Comparison}
The distributions of branch order and branch length across all three synthetic cell types closely match those observed in real cellular reconstructions. 
Similarly, tortuosity distributions for synthetic pyramidal and basket cells align well with their biological counterparts.

\begin{figure}[hbt!]
    \centering
    \includegraphics[scale=.85]{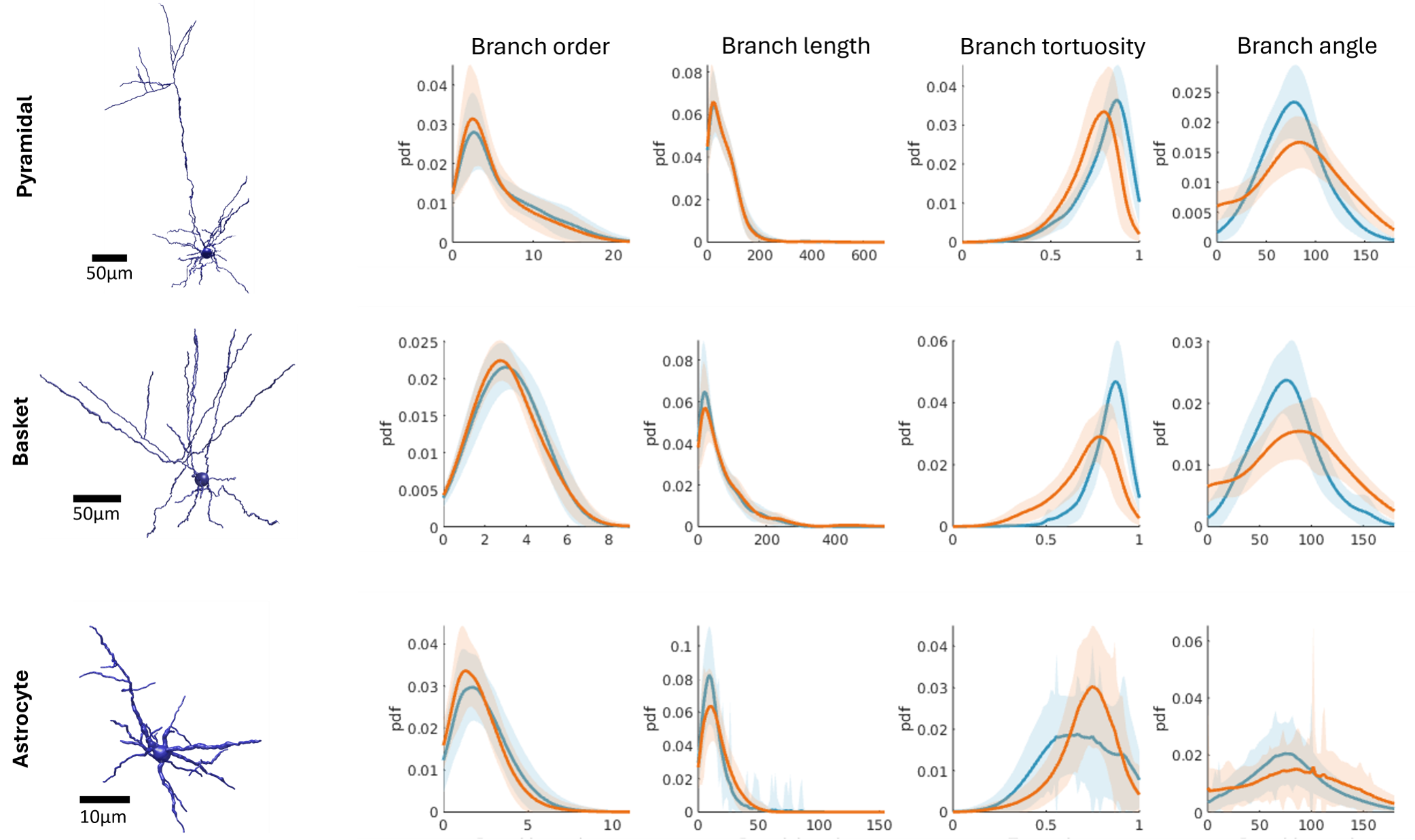}
    \caption{\textbf{Comparison of real (blue) and synthetic (orange) cellular characteristics}. Exemplar synthetic cells of type pyramidal, basket, and astrocyte, and distributions of branch order, branch length, tortuosity, and branch angle for both real and synthetic cells }
    \label{fig:sturcuralStats}
\end{figure}

Table \ref{tab:1} shows the comparison between the mean and standard deviation of structural features for real and ConCeG generated cells. There is good agreement between branch order, length, and angle, which is further supported when looking at the distributions of structural features seen in Figure \ref{fig:sturcuralStats}. The tortuosity is not as well replicated. This is expected since the tortuosity (differently from the branch order, length, and angle) was not controlled during growth. Nevertheless, the values (naturally arising from the growth) are still remarkably within realistic ranges \cite{aird2026decoding}.

These results indicate the ability of ConCeG to accurately replicate the cellular structures that informed their growth.

\begin{table}[]
\resizebox{\textwidth}{!}{
\begin{tabular}{ccc|ccccc}
Later                 & Cell type                  & Origin & $N$ & Branch order   & Branch length ($\mu m$) & $\tau$ & Branch angle \\ \hline
\multirow{4}{*}{2/3}  & \multirow{2}{*}{Pyramidal} & Real   & 12  & $3.9 \pm 0.91$ & $46 \pm 4.3$                         & $0.82\pm0.13$       & $70 \pm 8.2$ \\ \cline{3-3}
                      &                            & ConCeG & 180 & $3.8 \pm 0.90$ & $41 \pm 5.8$                         & $0.67\pm0.15$       & $73 \pm 9.9$ \\ \cline{2-8} 
                      & \multirow{2}{*}{Basket}    & Real   & 21  & $3.2 \pm 0.57$ & $48 \pm 14$                          & $0.82\pm0.12$       & $79 \pm 7.9$ \\ \cline{3-3}
                      &                            & ConCeG & 44  & $3.2 \pm 0.55$ & $42 \pm 13$                          & $0.67\pm0.16$       & $80 \pm 11$  \\ \hline
\multirow{4}{*}{4}    & \multirow{2}{*}{Pyramidal} & Real   & 73  & $3.8 \pm 0.63$ & $48 \pm 5.2$                         & $0.83\pm0.13$       & $72 \pm 7.1$ \\
                      &                            & ConCeG & 159 & $3.5 \pm 0.69$ & $41 \pm 8.7$                         & $0.68\pm0.15$       & $74 \pm 12$  \\ \cline{2-8} 
                      & \multirow{2}{*}{Basket}    & Real   & 12  & $3.2 \pm 1.1$  & $59 \pm 13$                          & $0.85\pm0.10$       & $74 \pm 11$  \\
                      &                            & ConCeG & 34  & $3.4 \pm 0.96$ & $42 \pm 18$                          & $0.67\pm0.15$       & $72 \pm 15$  \\ \hline
\multirow{4}{*}{5}    & \multirow{2}{*}{Pyramidal} & Real   & 60  & $5.7 \pm 2.5$  & $53 \pm 7.8$                         & $0.83\pm0.13$       & $74 \pm 6.6$ \\ \cline{3-3}
                      &                            & ConCeG & 221 & $4.9 \pm 1.9$  & $49 \pm 7.6$                         & $0.71\pm0.14$       & $75 \pm 9.0$ \\ \cline{2-8} 
                      & \multirow{2}{*}{Basket}    & Real   & 39  & $2.8 \pm 0.41$ & $59 \pm 15$                          & $0.86\pm0.10$       & $73 \pm 8.2$ \\ \cline{3-3}
                      &                            & ConCeG & 52  & $2.9 \pm 0.49$ & $58 \pm 17$                          & $0.72\pm0.14$       & $74 \pm 11$  \\ \hline
\multirow{4}{*}{6}    & \multirow{2}{*}{Pyramidal} & Real   & 33  & $5.7 \pm 1.5$  & $61 \pm 11$                          & $0.80\pm0.14$       & $77 \pm 8.7$ \\ \cline{3-3}
                      &                            & ConCeG & 213 & $5.2 \pm 1.6$  & $59 \pm 11$                          & $0.73\pm0.13$       & $78 \pm 9.8$ \\ \cline{2-3} \cline{5-8} 
                      & \multirow{2}{*}{Basket}    & Real   & 21  & $3.1 \pm 0.42$ & $60 \pm 15$                          & $0.82\pm0.11$       & $75 \pm 10$  \\ \cline{3-3}
                      &                            & ConCeG & 43  & $3.0 \pm 0.32$ & $70 \pm 16$                          & $0.71\pm0.15$       & $78 \pm 11$  \\ \hline
\multirow{2}{*}{N.A.} & \multirow{2}{*}{Astrocyte} & Real   & 845 & $2.2 \pm 0.77$ & $13 \pm 6.7$                         & $0.68\pm0.19$       & $75 \pm 16$  \\ \cline{3-3}
                      &                            & ConCeG & 590 & $1.9 \pm 0.62$ & $15 \pm 4.3$                         & $0.71\pm0.15$       & $78 \pm 24$  \\ \hline
\end{tabular}
}
\caption{\textbf{Comparison of real and ConCeG structural statistics}. Comparing, branch order, branch length, tortuosity, and branch angle. }
\label{tab:1}
\end{table}

\subsection{Intracellular Power Spectrum}
\begin{figure}[hbt!]
    \centering
    \includegraphics[scale=0.3]{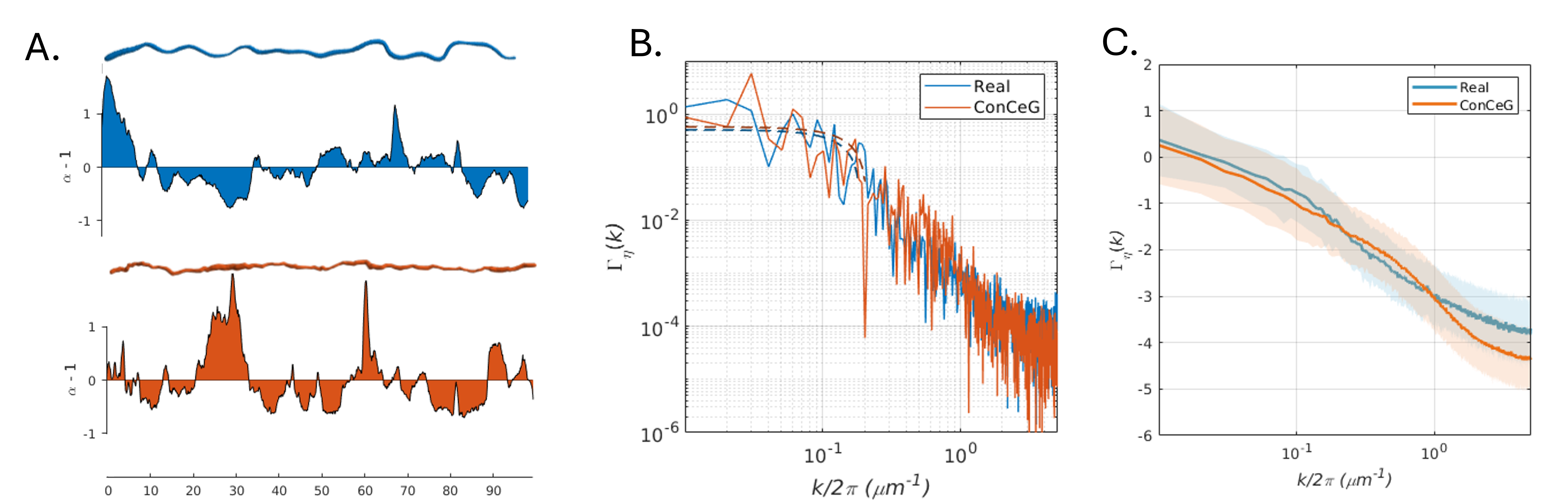}
    \caption{\textbf{Comparison of real and synthetic intra-cellular power spectrum}. A. 
    Shows the cross sectional area variation perpendicular to the length of the branch and corresponding power spectrum for real (orange) and ConCeG generated (blue) branches. B. Shows the computed power spectrum for the real and synthetic cross sectional area variation. C. Shows the mean power spectrum for all real and synthetic branches. }
    \label{fig:intraPower}
\end{figure}

The synthetic branches reproduce the similar mean power spectra observed in real branches, as shown in Figure \ref{fig:intraPower}C. 
Select individual branches do exhibit a plateau at low spatial frequencies consistent with short range disorder.
However, the mean power spectra for both real and synthetic branches do not exhibit a plateau at low spatial frequencies which is likely a due to tapering present in both real and ConCeG branches.

\subsection{Universal composition}
\begin{figure}[hbt!]
    \centering
    \includegraphics[scale=0.22]{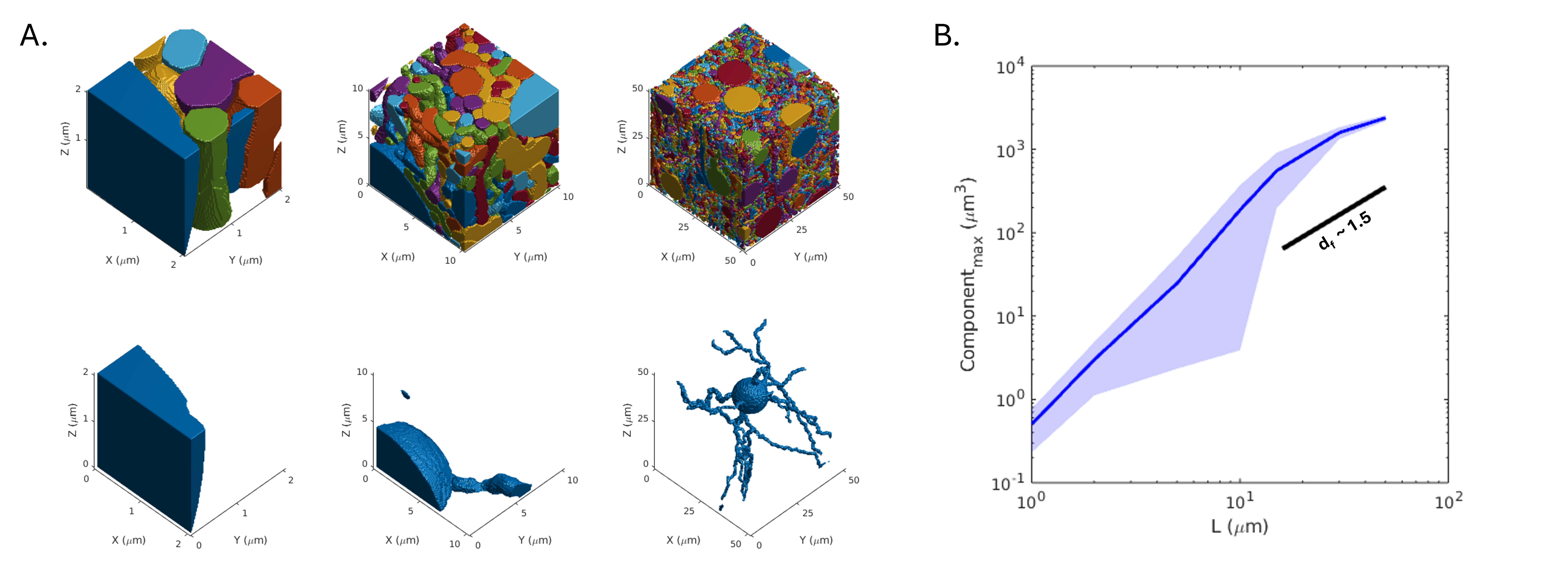}
    \caption{\textbf{Power law scaling of cellular composition}. A. shows a visualisation of sub voxel samples (2, 10 and 50 $\mu m$ side lengths) of the substrate (top) and the largest component contained with in the sub voxel (bottom). B. shows the relationship between largest component volume and sub voxel side length, the shaded area shows the standard deviation.  
     }
    \label{fig:Composition}
\end{figure}

As shown in Figure \ref{fig:Composition}B., the logarithm of the largest cellular component volume $(\mu m^{3})$ exhibits a clear linear relationship with the logarithm of the sampling length scale (L). 
This indicates a power-law scaling behaviour, consistent with fractal structure across length scales.

Furthermore, a linear fit in log–log space yields an estimated fractal dimension of approximately ~1.5, close to the fractal dimension of ~1.6 reported in \cite{Ansell2024}.
The agreement suggests that the synthetic substrates reproduce composition multi-scale statistics similar to those observed in real tissue.

\subsection{Extracellular Space}
\begin{figure}[hbt!]
    \centering
    \includegraphics[scale=.55]{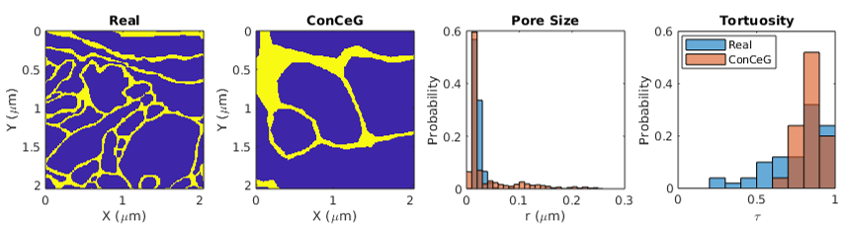}
    \caption{\textbf{Comparison of real and synthetic extra-cellular space}. Visual comparison of real and synthetic extracellular space, alongside the corresponding pore size and tortuosity distributions.}
    \label{fig:extraPower}
\end{figure}

The mean pore size in the real and ConCeG generated media shows close agreement as seen in Figure \ref{fig:extraPower}, indicating that the synthetic growth accurately reproduces the dominant length scale of the microstructure. 

However, differences are observed in the distribution tails. 
The ConCeG generated media exhibits an extended tail toward larger pore sizes that is not present in the real dataset. 
This indicates large pores are present in the ConCeG media that are not present in the comparison with em segmentation. 

The mean tortuosity is broadly similar between the real and ConCeG generated media, indicating that both datasets exhibit comparable overall path complexity and connectivity.

A reduction in tortuosity is observed in the ConCeG generated media relative to the real samples. This indicates that transport pathways in the generated structures tend to be more direct on average.

\subsection{Simulated signals}
The generated substrates are fully compatible with Monte Carlo diffusion simulations.
Diffusion-weighted signals were obtained across the full range of simulated b-values for both the complete substrate and the isolated intracellular compartment.

The simulated intracellular diffusion signal varied according to the underlying tissue composition. 
Notably, the substrate representing cortical layer 4 exhibited the greatest signal attenuation, consistent with its larger intracellular soma volume fraction. 
The decreased volume fraction of neurites in this layer results in a lower proportion of highly restricted intracellular space, producing a diffusion signal that differs from those of layers 3 and 5.

Furthermore, substrates representing layers 3 and 5 exhibit very similar signal attenuation. 
This reflects their comparable cellular composition, with both layers containing an intracellular soma volume fraction of approximately 25\%. 
As a result, the relative contributions of the soma and neurite compartments to the diffusion signal were also similar, resulting in similar diffusion-weighted signal attenuation.

The estimated SANDI parameters, as seen in Table \ref{tab:2}, show good agreement with the ground-truth values across the three substrates.
For the soma volume fraction, SANDI estimates were close to the ground truth for all substrates, with values of 28\%, 24\%, and 46\% compared with 25\%, 26\%, and 46\% respectively. 

The effective soma radius was consistently slightly overestimated by SANDI, with estimates of 7.8, 8.1, and 7.7 $\mu m$, compared with ground-truth values of 7.0, 7.2, and 6.9 $\mu m$, for layers 3, 4, and 5 respectively. 
Despite these discrepancies, the estimates reflect the relative variation between layers, with layer 4 exhibiting the largest soma volume fraction and effective soma radius. 

The SANDI parameter estimates for the full signal reproduced the overall trends observed for the intracellular parameters, with similar parameter estimates in Layers 3 and 5 and an increased soma volume fraction in Layer 4. 

However, the extracellular volume fraction was consistently underestimated by SANDI across all layers, with estimates of 16\%, 13\%, and 18\% in Layers 3, 4, and 5, respectively, compared with 31\%, 30\%, and 33\% ground truth values.

As with the intra cellular parameter estimation, the effective soma radius was consistently overestimated, with estimates of 7.7, 8.2, and 8.0 $\mu m$ in Layers 3, 4, and 5, respectively, compared with 7.0, 7.2, and 6.9 $\mu m$ for ground truth values.

These results demonstrate that the ConCeG substrates capture key features that influence diffusion MRI and reproduce trends in parameter estimates consistent with the ground-truth substrate geometries. 
This agreement supports the use of the synthetic substrates as a controlled framework for investigating the relationship between tissue microstructure, the resulting diffusion-weighted MR signal and the ability of biophysical models (e.g. SANDI) to capture the salient microstructural features.

\begin{figure}[hbt!]
    \centering
    \includegraphics[scale=1]{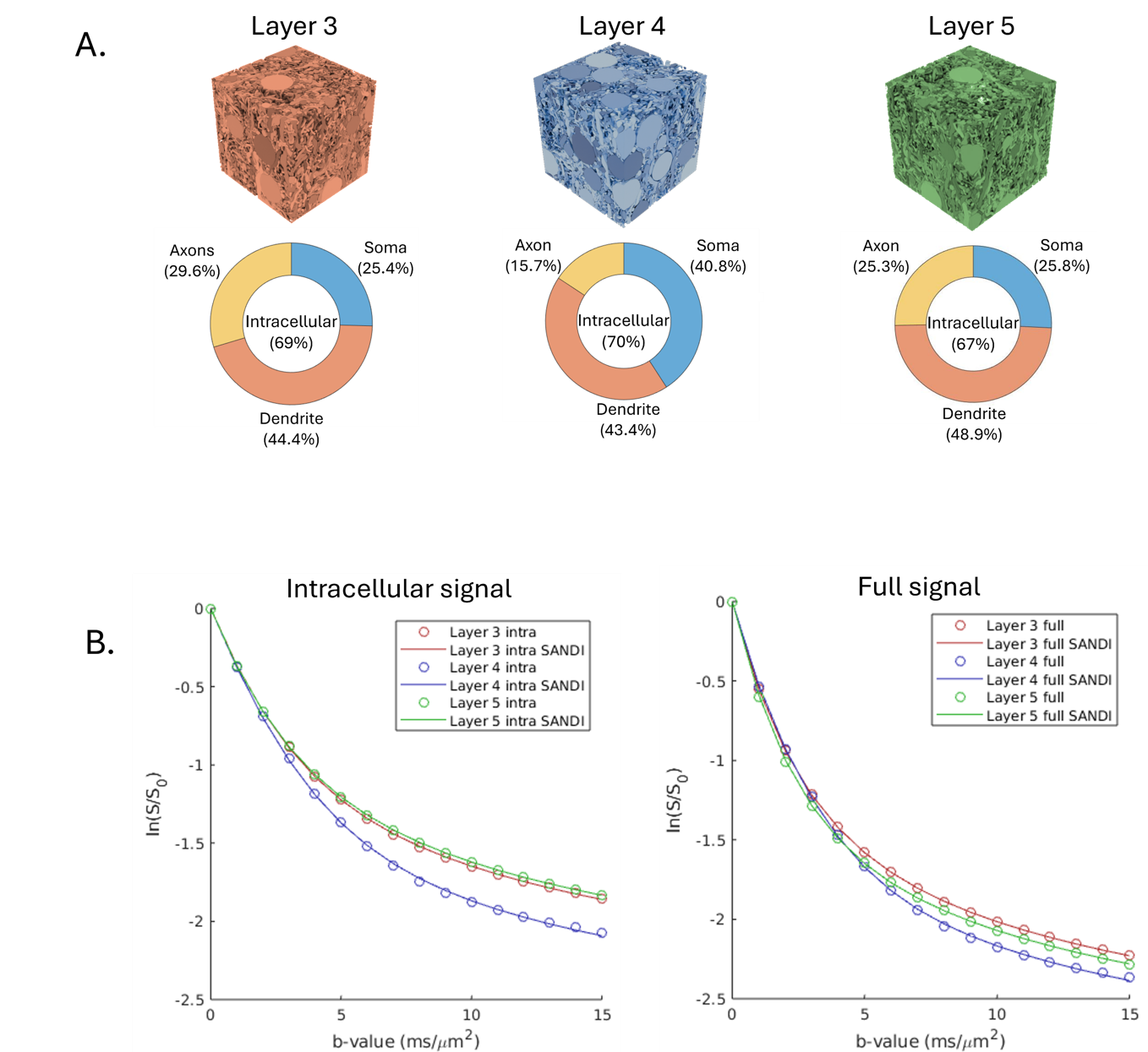}
    \caption{\textbf{Simulated signals}, showing both intracellular and complete signals for the three meshes generated from the cortical column seen in Figure \ref{fig:column}. }
    \label{fig:signal}
\end{figure}

\begin{table}[]
\resizebox{\textwidth}{!}{
\begin{tabular}{c|cccc|c|cccccccc|}
      & \multicolumn{4}{c|}{Intracellular}                                            &  & \multicolumn{8}{c|}{Full}                                                                                                                                       \\ \cline{2-5} \cline{7-14} 
      & \multicolumn{2}{c|}{$f_{Soma} (\%)$} & \multicolumn{2}{c|}{$R_{soma} (\mu m)$} &  & \multicolumn{2}{c|}{$f_{Neurite} (\%)$} & \multicolumn{2}{c|}{$f_{Soma} (\%)$} & \multicolumn{2}{c|}{$f_{Extra} (\%)$} & \multicolumn{2}{c|}{$R_{soma} (\mu m)$} \\ \cline{1-5} \cline{7-14} 
Layer & ConCeG  & \multicolumn{1}{c|}{SANDI} & ConCeG             & SANDI             &  & ConCeG   & \multicolumn{1}{c|}{SANDI}   & ConCeG  & \multicolumn{1}{c|}{SANDI} & ConCeG  & \multicolumn{1}{c|}{SANDI}  & ConCeG             & SANDI             \\ \cline{1-5} \cline{7-14} 
3     & 25      & \multicolumn{1}{c|}{28}    & 7.0                & 7.8               &  & 52       & \multicolumn{1}{c|}{64}      & 17      & \multicolumn{1}{c|}{20}    & 31      & \multicolumn{1}{c|}{16}     & 7.0                & 7.7               \\
4     & 41      & \multicolumn{1}{c|}{46}    & 7.2                & 7.8               &  & 42       & \multicolumn{1}{c|}{57}      & 28      & \multicolumn{1}{c|}{31}    & 30      & \multicolumn{1}{c|}{13}     & 7.2                & 8.2               \\
5     & 26      & \multicolumn{1}{c|}{24}    & 6.9                & 7.7               &  & 50       & \multicolumn{1}{c|}{63}      & 18      & \multicolumn{1}{c|}{19}    & 33      & \multicolumn{1}{c|}{18}     & 6.9                & 8.0              
\end{tabular}
}
\caption{\textbf{Comparison of ground truth values from the ConCeG substrates and estimated SANDI} parameters from intracellular and full simulated signals seen in Figure \ref{fig:signal}.}
\label{tab:2}
\end{table}

\section{Discussion}
The aim of this work was to introduce ConCeG: Contextual Cellular Growth, a computational modelling approach to generate realistic numerical phantoms for diffusion simulations by generating synthetic substrates informed by real cellular morphologies. 
The results demonstrate that the ConCeG framework successfully captures key structural and statistical features of biological tissue across different spatial scales.

On the cellular level, the good agreement between synthetic and real distributions of branch order, branch length, and branching angles indicates that the generative approach preserves the organisation of dendritic projections. 
The disparity observed in tortuosity should be interpreted with caution, as tortuosity was not explicitly constrained during the growth process.  Rather, it is notable that the synthetic cells reproduce a broadly comparable tortuosity distribution, and remaining within realistic values, despite this feature emerging indirectly from the growth rules rather than being imposed as a target constraint

Beyond  cell structure, the agreement in power spectra demonstrates that the synthetic substrates also replicate sub-cellular features. 
The ability of the intracellular space to reproduce the short-range disorder observed in real tissue indicates that the spatial arrangement of cellular components is realistically captured at microscopic scales.

Additionally, similar characteristic features are observed in the extracellular space, with pore size and tortuosity measurements showing broad agreement between EM-derived segmentation's and ConCeG generated tissue. This suggests that the contextual growth process is able to reproduce not only realistic intracellular morphologies but also aspects of the spatial organisation and packing of cells that give rise to extracellular geometry.

However, differences remain in the distribution tails. 
ConCeG generated tissue exhibits a greater proportion of larger pores than observed in the EM-derived tissue and a reduction in extracellular tortuosity. 
One possible explanation is tissue shrinkage associated with electron microscopy preparation, which has been shown to reduce extracellular volume fraction and alter extracellular geometry.
Alternatively, these discrepancies may reflect limitations of the current growth framework, including the limited node density of the growth network, simplified biological growth mechanisms that are only locally informed and indifferent to the global space, or the absence of additional microstructural features that contribute to extracellular organisation in vivo such as spines.

Despite these differences, the overall agreement in extracellular metrics suggests that ConCeG captures the dominant structural constraints governing extracellular architecture while identifying clear directions for future improvements of tissue packing.

The observed power-law scaling of component size suggests that the organisation of the synthetic tissue reproduces important compositional properties of real grey matter across spatial scales. 
This behaviour is consistent with observations in biological tissue and indicates that ConCeG captures not only realistic cellular morphologies but also properties of the emergent spatial organisation arising from the cellular packing.

This is further supported by the estimated fractal dimension of the synthetic tissue, which closely matches the values previously reported in human, mouse, and fly grey matter. 
Interestingly, this similarity emerges without enforcing fractal structure during growth, indicating that the combination of biologically informed morphology and contextual growth constraints is sufficient to recreate realistic large-scale organisation.

The generalisability of ConCeG should be considered in relation to the biological data used to inform substrate generation. The cortical column presented here is based on available mouse visual-cortex morphology and density information, and therefore represents a realistic, data-informed exemplar rather than a universal model of all GM. GM architecture varies across cortical areas, species, developmental stage, ageing, and pathology, with differences in laminar organisation, neuronal density, dendritic complexity, glial composition, and extracellular volume fraction. Consequently, applications to other brain regions or disease states will require region- and condition-specific input distributions, ideally derived from matched histological, electron microscopy, or high-resolution optical datasets. Because ConCeG is parameterised, it provides a framework for systematically perturbing cellular density, morphology, packing, and composition in controlled ways. This makes it well suited for testing hypotheses about how specific microstructural alterations, such as neuronal loss, dendritic regression, gliosis, or extracellular remodelling, may bias or drive diffusion MRI biomarkers in ageing and neurological disease.

Finally, it is important to note that, although agreement in morphometry and spatial power spectra provides evidence that ConCeG captures relevant structural length scales, this does not automatically guarantee realistic diffusion-weighted signals. 
The simulated diffusion-weighted signals and SANDI parameter estimates reported here do provide strong evidence that the simulated signals preserve relevant biophysical features and can be used as a tool to validate or challenge existing or new biophysical models. Worth noting that further validation of the time dependent diffusivity would be valuable and will be focus of future work. 

Overall, these findings suggest that the proposed framework provides a basis for generating realistic numerical phantoms for diffusion simulations. 

\subsection{Limitations}

Despite the promising results demonstrated by the ConCeG algorithm, several limitations should be acknowledged.

First, the current growth model relies on a finite static graph navigation network.
While this enables efficient synthesis of dense tissue, it may limit the ability to fully capture fine-scale morphological features, particularly for smaller cells such as astrocytes. Increasing node density of course improves fidelity but comes at a computational cost, future work can be done to implement a more dynamic network.

Secondly, whilst ConCeG is able to produce 100 $\mu m^{3}$ substrates with a packing density of 70\%, the meshing pipeline remains limited when compared to approaches like CACTUS \cite{Villarreal-Haro2023}, which is capable of generating 500 $\mu m^{3}$ high density meshes.
Fortunately, the SWC format of the ConCeG generated cells are output as is not dissimilar to the fibre structure employed in CACTUS so future work may be taken to develop a meshing pipeline with similar capabilities.
As the SWC structure of the ConCeG cells is similar to the fibre structure of CACTUS requires for meshing a similiar approach can potentially be taken in the future to allow for the generation of larger GM substrates.  

Additionally, validation is currently limited to comparisons against available morphological reconstructions and electron microscopy datasets, which are subject to sampling biases, reconstruction artifacts, and tissue preparation effects. 
As interest in GM continues to grow, larger and more diverse datasets are expected to become available
This should allow for more robust validation, and opportunities to refine and extend the proposed approach in future work.

Finally, the current implementation should also be viewed as a representation of selected grey-matter cellular backbones rather than a complete biological reconstruction of grey matter. Several structures that are likely to influence diffusion, relaxation, and exchange are not yet explicitly represented, including dendritic spines, synaptic boutons, intracellular organelles, nuclei, myelin fragments, vasculature, and perivascular spaces. This is particularly important because for example dendritic spines are expected to contribute substantially to surface-to-volume ratio, extracellular narrowing, water exchange, and diffusion-time-dependent signal behaviour in cortical GM \cite{chakwizira2025role, simsek2025role}. Future versions of ConCeG could therefore incorporate modular additions of spines, boutons, vessels. This would allow the same generative framework to be used not only to create anatomically realistic substrates, but also to test which biological features are detectable with realistic diffusion MRI acquisitions and which remain degenerate or poorly identifiable.

\section{Conclusion}

We presented ConCeG, a generative framework for constructing realistic grey matter substrates for diffusion MRI simulations. By combining morphology-informed growth with a constrained spatial network, the method produces dense multi-cellular environments that preserve key structural and topological features of neurons and glia.

Validation results show good agreement with biological reconstructions across multiple scales, including cellular morphology, intracellular structure, and emergent tissue organisation. The generated substrates are also compatible with Monte Carlo diffusion simulations.

Overall, ConCeG addresses a longstanding gap in diffusion MRI simulation by enabling the generation of biologically realistic, multicellular grey matter substrates with controllable composition and organisation. 
By bridging cellular morphology and tissue-scale architecture, it provides a powerful new platform for investigating the relationship between grey matter microstructure and diffusion MRI signals.

\section{Code availability}
The full Matlab, python, and blender implementaion will be available at \href{https://github.com/Charlie-Aird/ConCeG}{https://github.com/Charlie-Aird/ConCeG} upon paper publication.

\section{Acknowledgements}
This work, C.A.R., K.S., M.J. and M.P. are supported by the UKRI Future Leaders Fellowship MR/T020296/2 and UKRI1073. 
D.K.J. was supported by a Wellcome Trust Strategic Award (104943/Z/14/Z) and Wellcome Discovery Award (227882/Z/23/Z)
L.K . is supported by the Medical Research Council, UKRI (MR/Z504804/1).

\bibliographystyle{unsrt}
\bibliography{MyBib}

@article{novikov2019quantifying,
  title={Quantifying brain microstructure with diffusion MRI: Theory and parameter estimation},
  author={Novikov, Dmitry S and Fieremans, Els and Jespersen, Sune N and Kiselev, Valerij G},
  journal={NMR in Biomedicine},
  volume={32},
  number={4},
  pages={e3998},
  year={2019},
  publisher={Wiley Online Library}
}

@article{ianus2021mapping,
  title={Mapping complex cell morphology in the grey matter with double diffusion encoding MR: A simulation study},
  author={Ianus, Andrada and Alexander, Daniel C and Zhang, Hui and Palombo, Marco},
  journal={Neuroimage},
  volume={241},
  pages={118424},
  year={2021},
  publisher={Elsevier}
}

@article{lee2020time,
  title={A time-dependent diffusion MRI signature of axon caliber variations and beading},
  author={Lee, Hong-Hsi and Papaioannou, Antonios and Kim, Sung-Lyoung and Novikov, Dmitry S and Fieremans, Els},
  journal={Communications biology},
  volume={3},
  number={1},
  pages={354},
  year={2020},
  publisher={Nature Publishing Group UK London}
}

@article{palombo2019generative,
  title={A generative model of realistic brain cells with application to numerical simulation of the diffusion-weighted MR signal},
  author={Palombo, Marco and Alexander, Daniel C and Zhang, Hui},
  journal={NeuroImage},
  volume={188},
  pages={391--402},
  year={2019},
  publisher={Elsevier}
}

@article{kanari2018topological,
  title={A topological representation of branching neuronal morphologies},
  author={Kanari, Lida and D{\l}otko, Pawe{\l} and Scolamiero, Martina and Levi, Ran and Shillcock, Julian and Hess, Kathryn and Markram, Henry},
  journal={Neuroinformatics},
  volume={16},
  pages={3--13},
  year={2018},
  publisher={Springer}
}

@article{kanari2022computational,
  title={Computational synthesis of cortical dendritic morphologies},
  author={Kanari, Lida and Dictus, Hugo and Chalimourda, Athanassia and Arnaudon, Alexis and Van Geit, Werner and Coste, Benoit and Shillcock, Julian and Hess, Kathryn and Markram, Henry},
  journal={Cell Reports},
  volume={39},
  number={1},
  year={2022},
  publisher={Elsevier}
}

@book{callaghan1993principles,
  title={Principles of nuclear magnetic resonance microscopy},
  author={Callaghan, Paul T},
  year={1993},
  publisher={Clarendon press}
}

@article{ascoli2007neuromorpho,
  title={NeuroMorpho. Org: a central resource for neuronal morphologies},
  author={Ascoli, Giorgio A and Donohue, Duncan E and Halavi, Maryam},
  journal={Journal of Neuroscience},
  volume={27},
  number={35},
  pages={9247--9251},
  year={2007},
  publisher={Soc Neuroscience}
}

@article{zhang2012noddi,
  title={NODDI: practical in vivo neurite orientation dispersion and density imaging of the human brain},
  author={Zhang, Hui and Schneider, Torben and Wheeler-Kingshott, Claudia A and Alexander, Daniel C},
  journal={Neuroimage},
  volume={61},
  number={4},
  pages={1000--1016},
  year={2012},
  publisher={Elsevier}
}

@article{alexander2010orientationally,
  title={Orientationally invariant indices of axon diameter and density from diffusion MRI},
  author={Alexander, Daniel C and Hubbard, Penny L and Hall, Matt G and Moore, Elizabeth A and Ptito, Maurice and Parker, Geoff JM and Dyrby, Tim B},
  journal={Neuroimage},
  volume={52},
  number={4},
  pages={1374--1389},
  year={2010},
  publisher={Elsevier}
}

@article{veraart2020noninvasive,
  title={Noninvasive quantification of axon radii using diffusion MRI},
  author={Veraart, Jelle and Nunes, Daniel and Rudrapatna, Umesh and Fieremans, Els and Jones, Derek K and Novikov, Dmitry S and Shemesh, Noam},
  journal={elife},
  volume={9},
  pages={e49855},
  year={2020},
  publisher={eLife Sciences Publications, Ltd}
}

@article{jelescu2020challenges,
  title={Challenges for biophysical modeling of microstructure},
  author={Jelescu, Ileana O and Palombo, Marco and Bagnato, Francesca and Schilling, Kurt G},
  journal={Journal of Neuroscience Methods},
  volume={344},
  pages={108861},
  year={2020},
  publisher={Elsevier}
}

@article{abdellah2018neuromorphovis,
  title={NeuroMorphoVis: a collaborative framework for analysis and visualization of neuronal morphology skeletons reconstructed from microscopy stacks},
  author={Abdellah, Marwan and Hernando, Juan and Eilemann, Stefan and Lapere, Samuel and Antille, Nicolas and Markram, Henry and Sch{\"u}rmann, Felix},
  journal={Bioinformatics},
  volume={34},
  number={13},
  pages={i574--i582},
  year={2018},
  publisher={Oxford University Press}
}

@article{peng2021morphological,
  title={Morphological diversity of single neurons in molecularly defined cell types},
  author={Peng, Hanchuan and Xie, Peng and Liu, Lijuan and Kuang, Xiuli and Wang, Yimin and Qu, Lei and Gong, Hui and Jiang, Shengdian and Li, Anan and Ruan, Zongcai and others},
  journal={Nature},
  volume={598},
  number={7879},
  pages={174--181},
  year={2021},
  publisher={Nature Publishing Group UK London}
}

@article{chakwizira2025role,
  title={The role of dendritic spines in water exchange measurements with diffusion MRI: Double Diffusion Encoding and free-waveform MRI},
  author={Chakwizira, Arthur and {\c{S}}im{\c{s}}ek, Kadir and Szczepankiewicz, Filip and Palombo, Marco and Nilsson, Markus},
  journal={arXiv preprint arXiv:2504.21537},
  year={2025}
}

@techReport{Miller1984,
   author = {John P Miller and Gwen A Jacobs},
   journal = {J. exp. Biol},
   pages = {129-145},
   title = {RELATIONSHIPS BETWEEN NEURONAL STRUCTURE AND FUNCTION},
   volume = {112},
   year = {1984}
}

@misc{Surmeier2017,
   author = {D. James Surmeier and José A. Obeso and Glenda M. Halliday},
   doi = {10.1038/nrn.2016.178},
   issn = {14710048},
   issue = {2},
   journal = {Nature Reviews Neuroscience},
   month = {2},
   pages = {101-113},
   pmid = {28104909},
   publisher = {Nature Publishing Group},
   title = {Selective neuronal vulnerability in Parkinson disease},
   volume = {18},
   year = {2017}
}

@misc{Dickstein2013,
   author = {D. L. Dickstein and C. M. Weaver and J. I. Luebke and P. R. Hof},
   doi = {10.1016/j.neuroscience.2012.09.077},
   issn = {03064522},
   journal = {Neuroscience},
   month = {10},
   pages = {21-32},
   pmid = {23069756},
   title = {Dendritic spine changes associated with normal aging},
   volume = {251},
   year = {2013}
}

@article{Hall2009,
   author = {Matt G. Hall and Daniel C. Alexander},
   doi = {10.1109/TMI.2009.2015756},
   issn = {1558254X},
   issue = {9},
   journal = {IEEE Transactions on Medical Imaging},
   pages = {1354-1364},
   pmid = {19273001},
   title = {Convergence and Parameter Choice for Monte-Carlo Simulations of Diffusion MRI},
   volume = {28},
   year = {2009}
}

@misc{Lee2022,
   author = {Jiseon Lee and Hee Jin Kim},
   doi = {10.3389/fnagi.2022.931536},
   issn = {16634365},
   journal = {Frontiers in Aging Neuroscience},
   month = {6},
   publisher = {Frontiers Media S.A.},
   title = {Normal Aging Induces Changes in the Brain and Neurodegeneration Progress: Review of the Structural, Biochemical, Metabolic, Cellular, and Molecular Changes},
   volume = {14},
   year = {2022}
}

@article{Baloyannis2009,
   author = {S. J. Baloyannis},
   doi = {10.1016/j.jns.2009.02.370},
   issn = {0022510X},
   issue = {1-2},
   journal = {Journal of the Neurological Sciences},
   month = {8},
   pages = {153-157},
   pmid = {19296966},
   title = {Dendritic pathology in Alzheimer's disease},
   volume = {283},
   year = {2009}
}

@article{Nilsson2009,
   author = {Markus Nilsson and Jimmy Lätt and Emil Nordh and Ronnie Wirestam and Freddy Ståhlberg and Sara Brockstedt},
   doi = {10.1016/j.mri.2008.06.003},
   issn = {0730725X},
   issue = {2},
   journal = {Magnetic Resonance Imaging},
   month = {2},
   pages = {176-187},
   pmid = {18657924},
   title = {On the effects of a varied diffusion time in vivo: is the diffusion in white matter restricted?},
   volume = {27},
   year = {2009}
}

@article{Nassif2022,
   author = {Caren Nassif and Allegra Kawles and Ivan Ayala and Grace Minogue and Nathan P. Gill and Robert A. Shepard and Antonia Zouridakis and Rachel Keszycki and Hui Zhang and Qinwen Mao and Margaret E. Flanagan and Eileen H. Bigio and M. Marsel Mesulam and Emily Rogalski and Changiz Geula and Tamar Gefen},
   doi = {10.1523/JNEUROSCI.0679-22.2022},
   issn = {15292401},
   issue = {45},
   journal = {Journal of Neuroscience},
   month = {11},
   pages = {8587-8594},
   pmid = {36180225},
   publisher = {Society for Neuroscience},
   title = {Integrity of Neuronal Size in the Entorhinal Cortex Is a Biological Substrate of Exceptional Cognitive Aging},
   volume = {42},
   year = {2022}
}

@article{Kweon2017,
   author = {Jung Hyun Kweon and Sunhong Kim and Sung Bae Lee},
   doi = {10.5483/BMBRep.2017.50.1.131},
   issn = {1976670X},
   issue = {1},
   journal = {BMB Reports},
   pages = {5-11},
   pmid = {27502014},
   publisher = {The Biochemical Society of the Republic of Korea},
   title = {The cellular basis of dendrite pathology in neurodegenerative diseases},
   volume = {50},
   year = {2017}
}

@article{Abdollahzadeh2025,
   author = {Ali Abdollahzadeh and Ricardo Coronado-Leija and Hong Hsi Lee and Alejandra Sierra and Els Fieremans and Dmitry S. Novikov},
   doi = {10.1038/s41467-025-64793-1},
   issn = {20411723},
   issue = {1},
   journal = {Nature Communications },
   month = {12},
   pmid = {41198676},
   publisher = {Nature Research},
   title = {Scattering approach to diffusion quantifies axonal damage in brain injury},
   volume = {16},
   year = {2025}
}

@article{Fieremans2010,
   author = {Els Fieremans and Dmitry S. Novikov and Jens H. Jensen and Joseph A. Helpern},
   doi = {10.1002/nbm.1577},
   issn = {09523480},
   issue = {7},
   journal = {NMR in Biomedicine},
   month = {8},
   pages = {711-724},
   pmid = {20882537},
   title = {Monte Carlo study of a two-compartment exchange model of diffusion},
   volume = {23},
   year = {2010}
}

@article{Winther2024,
   author = {S. Winther and H. Lundell and J. Rafael-Patiño and M. Andersson and J. P. Thiran and T. B. Dyrby},
   doi = {10.1038/s41598-024-79043-5},
   issn = {20452322},
   issue = {1},
   journal = {Scientific Reports},
   month = {12},
   pmid = {39609481},
   publisher = {Nature Research},
   title = {Susceptibility-induced internal gradients reveal axon morphology and cause anisotropic effects in the diffusion-weighted MRI signal},
   volume = {14},
   year = {2024}
}

@article{Bihan1995,
   author = {Denis Le Bihan},
   doi = {10.1002/nbm.1940080711},
   issn = {10991492},
   issue = {7},
   journal = {NMR in Biomedicine},
   pages = {375-386},
   pmid = {8739274},
   title = {Molecular diffusion, tissue microdynamics and microstructure},
   volume = {8},
   year = {1995}
}

@article{Ginsburger2018,
   author = {Kévin Ginsburger and Fabrice Poupon and Justine Beaujoin and Delphine Estournet and Felix Matuschke and Jean François Mangin and Markus Axer and Cyril Poupon},
   doi = {10.3389/fphy.2018.00012},
   issn = {2296424X},
   issue = {FEB},
   journal = {Frontiers in Physics},
   month = {2},
   publisher = {Frontiers Media SA},
   title = {Improving the realism of white matter numerical phantoms: A step toward a better understanding of the influence of structural disorders in diffusion MRI},
   volume = {5},
   year = {2018}
}

@article{Lee2020,
   author = {Hong Hsi Lee and Antonios Papaioannou and Dmitry S. Novikov and Els Fieremans},
   doi = {10.1016/j.neuroimage.2020.117054},
   issn = {10959572},
   journal = {NeuroImage},
   month = {11},
   pmid = {32585341},
   publisher = {Academic Press Inc.},
   title = {In vivo observation and biophysical interpretation of time-dependent diffusion in human cortical gray matter},
   volume = {222},
   year = {2020}
}

@article{Fang2020,
   author = {Chengran Fang and Van Dang Nguyen and Demian Wassermann and Jing Rebecca Li},
   doi = {10.1016/j.neuroimage.2020.117198},
   issn = {10959572},
   journal = {NeuroImage},
   month = {11},
   pmid = {32730957},
   publisher = {Academic Press Inc.},
   title = {Diffusion MRI simulation of realistic neurons with SpinDoctor and the Neuron Module},
   volume = {222},
   year = {2020}
}

@article{Ansell2024,
   author = {Helen S. Ansell and István A. Kovács},
   doi = {10.1038/s42005-024-01665-y},
   issn = {23993650},
   issue = {1},
   journal = {Communications Physics},
   month = {12},
   publisher = {Nature Research},
   title = {Unveiling universal aspects of the cellular anatomy of the brain},
   volume = {7},
   year = {2024}
}

@article{Nilsson2010,
   author = {M. Nilsson and E. Alerstam and R. Wirestam and F. Ståhlberg and S. Brockstedt and J. Lätt},
   doi = {10.1016/j.jmr.2010.06.002},
   issn = {10907807},
   issue = {1},
   journal = {Journal of Magnetic Resonance},
   month = {9},
   pages = {59-67},
   pmid = {20594881},
   title = {Evaluating the accuracy and precision of a two-compartment Kärger model using Monte Carlo simulations},
   volume = {206},
   year = {2010}
}

@misc{Dickstein2007,
   author = {Dara L. Dickstein and Doron Kabaso and Anne B. Rocher and Jennifer I. Luebke and Susan L. Wearne and Patrick R. Hof},
   doi = {10.1111/j.1474-9726.2007.00289.x},
   issn = {14749718},
   issue = {3},
   journal = {Aging Cell},
   month = {6},
   pages = {275-284},
   pmid = {17465981},
   title = {Changes in the structural complexity of the aged brain},
   volume = {6},
   year = {2007}
}

@article{Rafael-Patino2020,
   author = {Jonathan Rafael-Patino and David Romascano and Alonso Ramirez-Manzanares and Erick Jorge Canales-Rodríguez and Gabriel Girard and Jean Philippe Thiran},
   doi = {10.3389/fninf.2020.00008},
   issn = {16625196},
   journal = {Frontiers in Neuroinformatics},
   month = {3},
   publisher = {Frontiers Media S.A.},
   title = {Robust Monte-Carlo Simulations in Diffusion-MRI: Effect of the Substrate Complexity and Parameter Choice on the Reproducibility of Results},
   volume = {14},
   year = {2020}
}

@article{Kerkel2020,
   author = {Leevi Kerkelä and Fabio Nery and Matt Hall and Chris Clark},
   doi = {10.21105/joss.02527},
   issue = {52},
   journal = {Journal of Open Source Software},
   month = {8},
   pages = {2527},
   publisher = {The Open Journal},
   title = {Disimpy: A massively parallel Monte Carlo simulator for generating diffusion-weighted MRI data in Python},
   volume = {5},
   year = {2020}
}

@article{Villarreal-Haro2023,
   author = {Juan Luis Villarreal-Haro and Remy Gardier and Erick J. Canales-Rodríguez and Elda Fischi-Gomez and Gabriel Girard and Jean Philippe Thiran and Jonathan Rafael-Patiño},
   doi = {10.3389/fninf.2023.1208073},
   issn = {16625196},
   journal = {Frontiers in Neuroinformatics},
   publisher = {Frontiers Media SA},
   title = {CACTUS: a computational framework for generating realistic white matter microstructure substrates},
   volume = {17},
   year = {2023}
}

@article{Tsai2009,
   author = {Philbert S. Tsai and John P. Kaufhold and Pablo Blinder and Beth Friedman and Patrick J. Drew and Harvey J. Karten and Patrick D. Lyden and David Kleinfeld},
   doi = {10.1523/JNEUROSCI.3287-09.2009},
   issn = {02706474},
   issue = {46},
   journal = {Journal of Neuroscience},
   month = {11},
   pages = {14553-14570},
   pmid = {19923289},
   title = {Correlations of neuronal and microvascular densities in murine cortex revealed by direct counting and colocalization of nuclei and vessels},
   volume = {29},
   year = {2009}
}

@article{Ginsburger2019,
   author = {Kévin Ginsburger and Felix Matuschke and Fabrice Poupon and Jean François Mangin and Markus Axer and Cyril Poupon},
   doi = {10.1016/j.neuroimage.2019.02.055},
   issn = {10959572},
   journal = {NeuroImage},
   month = {6},
   pages = {10-24},
   pmid = {30849528},
   publisher = {Academic Press Inc.},
   title = {MEDUSA: A GPU-based tool to create realistic phantoms of the brain microstructure using tiny spheres},
   volume = {193},
   year = {2019}
}

@misc{Assaf2008,
   author = {Yaniv Assaf and Ofer Pasternak},
   doi = {10.1007/s12031-007-0029-0},
   issn = {08958696},
   issue = {1},
   journal = {Journal of Molecular Neuroscience},
   month = {1},
   pages = {51-61},
   pmid = {18157658},
   title = {Diffusion tensor imaging (DTI)-based white matter mapping in brain research: A review},
   volume = {34},
   year = {2008}
}

@misc{Ascoli2007,
   author = {Giorgio A. Ascoli and Duncan E. Donohue and Maryam Halavi},
   doi = {10.1523/JNEUROSCI.2055-07.2007},
   issn = {02706474},
   issue = {35},
   journal = {Journal of Neuroscience},
   month = {8},
   pages = {9247-9251},
   pmid = {17728438},
   title = {NeuroMorpho.Org: A central resource for neuronal morphologies},
   volume = {27},
   year = {2007}
}

@article{Gouwens2019,
   author = {Nathan W. Gouwens and Staci A. Sorensen and Jim Berg and Changkyu Lee and Tim Jarsky and Jonathan Ting and Susan M. Sunkin and David Feng and Costas A. Anastassiou and Eliza Barkan and Kris Bickley and Nicole Blesie and Thomas Braun and Krissy Brouner and Agata Budzillo and Shiella Caldejon and Tamara Casper and Dan Castelli and Peter Chong and Kirsten Crichton and Christine Cuhaciyan and Tanya L. Daigle and Rachel Dalley and Nick Dee and Tsega Desta and Song Lin Ding and Samuel Dingman and Alyse Doperalski and Nadezhda Dotson and Tom Egdorf and Michael Fisher and Rebecca A. de Frates and Emma Garren and Marissa Garwood and Amanda Gary and Nathalie Gaudreault and Keith Godfrey and Melissa Gorham and Hong Gu and Caroline Habel and Kristen Hadley and James Harrington and Julie A. Harris and Alex Henry and Di Jon Hill and Sam Josephsen and Sara Kebede and Lisa Kim and Matthew Kroll and Brian Lee and Tracy Lemon and Katherine E. Link and Xiaoxiao Liu and Brian Long and Rusty Mann and Medea McGraw and Stefan Mihalas and Alice Mukora and Gabe J. Murphy and Lindsay Ng and Kiet Ngo and Thuc Nghi Nguyen and Philip R. Nicovich and Aaron Oldre and Daniel Park and Sheana Parry and Jed Perkins and Lydia Potekhina and David Reid and Miranda Robertson and David Sandman and Martin Schroedter and Cliff Slaughterbeck and Gilberto Soler-Llavina and Josef Sulc and Aaron Szafer and Bosiljka Tasic and Naz Taskin and Corinne Teeter and Nivretta Thatra and Herman Tung and Wayne Wakeman and Grace Williams and Rob Young and Zhi Zhou and Colin Farrell and Hanchuan Peng and Michael J. Hawrylycz and Ed Lein and Lydia Ng and Anton Arkhipov and Amy Bernard and John W. Phillips and Hongkui Zeng and Christof Koch},
   doi = {10.1038/s41593-019-0417-0},
   issn = {15461726},
   issue = {7},
   journal = {Nature Neuroscience},
   month = {7},
   pages = {1182-1195},
   pmid = {31209381},
   publisher = {Nature Publishing Group},
   title = {Classification of electrophysiological and morphological neuron types in the mouse visual cortex},
   volume = {22},
   year = {2019}
}

@article{Nilsson2012,
   author = {Markus Nilsson and Jimmy Lätt and Freddy Ståhlberg and Danielle van Westen and Håkan Hagslätt},
   doi = {10.1002/nbm.1795},
   issn = {09523480},
   issue = {5},
   journal = {NMR in Biomedicine},
   month = {5},
   pages = {795-805},
   pmid = {22020832},
   title = {The importance of axonal undulation in diffusion MR measurements: A Monte Carlo simulation study},
   volume = {25},
   year = {2012}
}

@article{Brabec2020,
   author = {Jan Brabec and Samo Lasič and Markus Nilsson},
   doi = {10.1002/nbm.4187},
   issn = {10991492},
   issue = {3},
   journal = {NMR in Biomedicine},
   month = {3},
   pmid = {31868995},
   publisher = {John Wiley and Sons Ltd},
   title = {Time-dependent diffusion in undulating thin fibers: Impact on axon diameter estimation},
   volume = {33},
   year = {2020}
}

@misc{Zeng2017,
   author = {Hongkui Zeng and Joshua R. Sanes},
   doi = {10.1038/nrn.2017.85},
   issn = {14710048},
   issue = {9},
   journal = {Nature Reviews Neuroscience},
   month = {8},
   pages = {530-546},
   pmid = {28775344},
   publisher = {Nature Publishing Group},
   title = {Neuronal cell-type classification: Challenges, opportunities and the path forward},
   volume = {18},
   year = {2017}
}

@article{Callaghan2020,
   author = {Ross Callaghan and Daniel C. Alexander and Marco Palombo and Hui Zhang},
   doi = {10.1016/j.neuroimage.2020.117107},
   issn = {10959572},
   journal = {NeuroImage},
   month = {10},
   pmid = {32622984},
   publisher = {Academic Press Inc.},
   title = {ConFiG: Contextual Fibre Growth to generate realistic axonal packing for diffusion MRI simulation},
   volume = {220},
   year = {2020}
}

@misc{Alexander2019,
   author = {Daniel C. Alexander and Tim B. Dyrby and Markus Nilsson and Hui Zhang},
   doi = {10.1002/nbm.3841},
   issn = {10991492},
   issue = {4},
   journal = {NMR in Biomedicine},
   month = {4},
   pmid = {29193413},
   publisher = {John Wiley and Sons Ltd},
   title = {Imaging brain microstructure with diffusion MRI: practicality and applications},
   volume = {32},
   year = {2019}
}

@misc{Jelescu2017,
   author = {Ileana O. Jelescu and Matthew D. Budde},
   doi = {10.3389/fphy.2017.00061},
   issn = {2296424X},
   issue = {NOV},
   journal = {Frontiers in Physics},
   month = {11},
   publisher = {Frontiers Media SA},
   title = {Design and validation of diffusion MRI models of white matter},
   volume = {5},
   year = {2017}
}

@book{levitan2002neuron,
  title={The neuron: cell and molecular biology},
  author={Levitan, Irwin B and Kaczmarek, Leonard K},
  year={2002},
  publisher={oxford university press}
}

@article{ross2011huntington,
  title={Huntington's disease: from molecular pathogenesis to clinical treatment},
  author={Ross, Christopher A and Tabrizi, Sarah J},
  journal={The Lancet Neurology},
  volume={10},
  number={1},
  pages={83--98},
  year={2011},
  publisher={Elsevier}
}

@article{castano2025architecture,
  title={Architecture and cellular composition of focal cortical dysplasia type II: qualitative review of histological studies},
  author={Casta{\~n}o-Mart{\'\i}n, Reyes and Metais, Alice and Ciura, Sorana and Blauwblomme, Thomas},
  journal={Frontiers in Cellular Neuroscience},
  volume={19},
  pages={1708220},
  year={2025},
  publisher={Frontiers Media SA}
}

@article{stejskal1965spin,
  title={Spin diffusion measurements: spin echoes in the presence of a time-dependent field gradient},
  author={Stejskal, Edward O and Tanner, John E},
  journal={The journal of chemical physics},
  volume={42},
  number={1},
  pages={288--292},
  year={1965},
  publisher={American Institute of Physics}
}

@article{budde2010neurite,
  title={Neurite beading is sufficient to decrease the apparent diffusion coefficient after ischemic stroke},
  author={Budde, Matthew D and Frank, Joseph A},
  journal={Proceedings of the National Academy of Sciences},
  volume={107},
  number={32},
  pages={14472--14477},
  year={2010},
  publisher={National Academy of Sciences}
}

@article{landman2010complex,
  title={Complex geometric models of diffusion and relaxation in healthy and damaged white matter},
  author={Landman, Bennett A and Farrell, Jonathan AD and Smith, Seth A and Reich, Daniel S and Calabresi, Peter A and Van Zijl, Peter CM},
  journal={NMR in Biomedicine: An International Journal Devoted to the Development and Application of Magnetic Resonance In vivo},
  volume={23},
  number={2},
  pages={152--162},
  year={2010},
  publisher={Wiley Online Library}
}

@article{nguyen2026caterpillar,
  title={CATERPillar: a flexible framework for generating white matter numerical substrates with incorporated glial cells},
  author={Nguyen-Duc, Jasmine and Brammerloh, Malte and Cherchali, Melina and De Riedmatten, In{\`e}s and P{\'e}rot, Jean-Baptiste and Rafael-Pati{\~n}o, Jonathan and Jelescu, Ileana O},
  journal={Medical Image Analysis},
  pages={103946},
  year={2026},
  publisher={Elsevier}
}

@article{abdollahzadeh2025scattering,
  title={Scattering approach to diffusion quantifies axonal damage in brain injury},
  author={Abdollahzadeh, Ali and Coronado-Leija, Ricardo and Lee, Hong-Hsi and Sierra, Alejandra and Fieremans, Els and Novikov, Dmitry S},
  journal={Nature Communications},
  volume={16},
  number={1},
  pages={9808},
  year={2025},
  publisher={Nature Publishing Group UK London}
}

@article{lee2021realistic,
  title={Realistic Microstructure Simulator (RMS): Monte Carlo simulations of diffusion in three-dimensional cell segmentations of microscopy images},
  author={Lee, Hong-Hsi and Fieremans, Els and Novikov, Dmitry S},
  journal={Journal of neuroscience methods},
  volume={350},
  pages={109018},
  year={2021},
  publisher={Elsevier}
}

@article{kiselev2026does,
  title={What Does FEXI Measure in Neurons?},
  author={Kiselev, Valerij G and Li, Jing-Rebecca},
  journal={arXiv preprint arXiv:2601.20657},
  year={2026}
}

@article{andersson2020axon,
  title={Axon morphology is modulated by the local environment and impacts the noninvasive investigation of its structure--function relationship},
  author={Andersson, Mariam and Kjer, Hans Martin and Rafael-Patino, Jonathan and Pacureanu, Alexandra and Pakkenberg, Bente and Thiran, Jean-Philippe and Ptito, Maurice and Bech, Martin and Bjorholm Dahl, Anders and Andersen Dahl, Vedrana and others},
  journal={Proceedings of the National Academy of Sciences},
  volume={117},
  number={52},
  pages={33649--33659},
  year={2020},
  publisher={National Academy of Sciences}
}

@article{microns2025functional,
  title={Functional connectomics spanning multiple areas of mouse visual cortex},
  journal={Nature},
  volume={640},
  number={8058},
  pages={435--447},
  year={2025},
  publisher={Nature Publishing Group UK London}
}

@dataset{MICrONS2021,
  author       = {{MICrONS Consortium}},
  title        = {MICrONS: Machine Intelligence from Cortical Networks},
  year         = {2021},
  month        = dec,
  publisher    = {Zenodo},
  doi          = {10.5281/zenodo.5760218},
  url          = {https://doi.org/10.5281/zenodo.5760218}
}

@article{coelho2022reproducibility,
  title={Reproducibility of the Standard Model of diffusion in white matter on clinical MRI systems},
  author={Coelho, Santiago and Baete, Steven H and Lemberskiy, Gregory and Ades-Aron, Benjamin and Barrol, Genevieve and Veraart, Jelle and Novikov, Dmitry S and Fieremans, Els},
  journal={NeuroImage},
  volume={257},
  pages={119290},
  year={2022},
  publisher={Elsevier}
}

@article{simsek2025role,
  title={The role of dendritic spines in water exchange measurements with diffusion mri: Time-dependent single diffusion encoding mri},
  author={{\c{S}}im{\c{s}}ek, Kadir and Chakwizira, Arthur and Nilsson, Markus and Palombo, Marco},
  journal={arXiv preprint arXiv:2506.18229},
  year={2025}
}

@article{aird2026decoding,
  title={Decoding gray matter, large-scale analysis of brain cell morphometry to inform microstructural modeling of diffusion MR signals},
  author={Aird-Rossiter, Charlie and Zhang, Hui and Alexander, Daniel C and Jones, Derek K and Palombo, Marco},
  journal={Communications Biology},
  year={2026},
  publisher={Nature Publishing Group UK London}
}

@incollection{dijkstra2022note,
  title={A note on two problems in connexion with graphs},
  author={Dijkstra, Edsger W},
  booktitle={Edsger Wybe Dijkstra: his life, work, and legacy},
  pages={287--290},
  year={2022}
}

\end{document}